\pdfoutput=1
\documentclass[journal]{IEEEtran}

\usepackage{tabularx,amsfonts,amssymb,amsmath}
\usepackage[dvips]{color}
\usepackage{epsfig}
\usepackage{mathrsfs}
\usepackage{subfigure}
\usepackage{calc,ifthen}
\usepackage{multirow}
\usepackage{bm}
\usepackage{float}
\usepackage{pstricks}
\usepackage{amsthm}
\usepackage{color}
\usepackage{empheq}
\usepackage{setspace}

\usepackage{cite}
\usepackage{psfrag}
\usepackage{pstool}
\usepackage{changes}
\usepackage{mathtools}
\usepackage{cuted}
\usepackage{gensymb}
\usepackage{color, colortbl}

\newcommand{\mynote}[1]{}
\newcommand{\newnote}[1]{}
\newcommand{\cancelled}[1]

\title{Mu-synthesis PID Control of Full-Car with Parallel Active Link Suspension Under Variable Payload}
\author{Zilin Feng, Min Yu, Simos A. Evangelou, Imad M Jaimoukha and Daniele Dini
\thanks{Zilin Feng ({\tt\small
zilin.feng17@imperial.ac.uk}), S. A. Evangelou {\tt\small (s.evangelou@imperial.ac.uk)} and Imad M Jaimoukha{\tt\small (i.jaimouka@imperial.ac.uk)} are with the Dept. of Electrical and Electronic at Imperial College London, UK .}
\thanks{Min Yu {\tt\small
(m.yu14@imperial.ac.uk)} and Daniele Dini {\tt\small
(d.dini@imperial.ac.uk)} are with the Dept. of Mechanical Engineering at Imperial College London, UK.}}
\begin{document}
\bstctlcite{IEEEexample:BSTcontrol}

\maketitle

\begin{abstract}
This paper presents a combined $\mu$-synthesis PID control scheme, employing a frequency separation paradigm, for a recently proposed novel active suspension, the Parallel Active Link Suspension (PALS). The developed $\mu$-synthesis control scheme is superior to the conventional $H_{\infty}$ control, previously designed for the PALS, in terms of ride comfort and road holding (higher frequency dynamics), with important realistic uncertainties, such as in vehicle payload, taken into account. The developed PID control method is applied to guarantee good chassis attitude control capabilities and minimization of pitch and roll motions (low frequency dynamics).
A multi-objective control method, which merges the aforementioned PID and $\mu$-synthesis-based controls is further introduced to achieve simultaneously the low frequency mitigation of attitude motions and the high frequency vibration suppression of the vehicle. A seven-degree-of-freedom Sport Utility Vehicle (SUV) full car model with PALS, is employed in this work to test the synthesized controller 
by nonlinear simulations with different ISO-defined road events and variable vehicle payload. The results demonstrate the control scheme's significant robustness 
and performance, as compared to the conventional passive suspension as well as the actively controlled PALS by conventional $H_{\infty}$ control, achieved for a wide range of vehicle payload considered in the investigation.

\end{abstract}
\vspace{-2mm}

\section{Introduction}
\IEEEPARstart{T}{he} main purposes of the suspension system in a car are to enhance ride quality for passengers and to allow the tires to maintain contact with the road to improve handling, which is achieved by controlling the vertical, pitch and roll motions of the car body and vertical motion of the tires. 
Over the past decades, passive, semi-active and active suspension systems have been presented to improve the vehicle ride quality\cite{sharp1987road}. Passive suspension is widely used in automotive applications due to low cost and simple implementation. Nevertheless, it can only passively adapt to the road profile and thus its performance is limited by the trade-off that exists between ride comfort and handling. Semi-active suspension has the same mechanical layout as the passive one, whereas the viscous damping coefficient of the shock absorber can be actively adjusted by a control unit. It provides less compromise in terms of ride comfort and handling over passive suspension and significantly less power consumption as compared to active suspension, however, it still lacks behind active suspension in terms of performance. 
Active suspensions
offer the best performance in relation to ride comfort and road holding, while their weaknesses are associated with weight, size, power requirements and complexity, with many different designs currently being employed in the industry including hydraulic and pneumatic solutions\cite{Aranaphdthesis}. 

Many active suspension control approaches have been proposed in the last few decades, such as PID (Proportional Integral Derivative) control\cite{talib2013self,ahmed2015pid,khodadadi2018self}, back-stepping control\cite{sun2012adaptive,pang2019adaptive}, sliding mode control\cite{deshpande2014disturbance,liu2020adaptive}, $H_{\infty}$ control\cite{wang2017robust,jing2014output}, MPC (Model predictive control)\cite{gohrle2012active,gohrle2013design}, fuzzy control\cite{wang2012hierarchical}, and constrained adaptive robust control\cite{sun2014vibration}. In \cite{pang2019adaptive}, the backstepping method is introduced as one of the significant approaches widely employed for suspension control. The constrained adaptive robust control scheme is proposed in \cite{sun2014vibration} for a quarter car active suspension system in the presence of parametric uncertainties and actuator saturation to stabilize the attitude of the vehicle. In \cite{wang2017robust}, the non-fragile $H_{\infty}$ static output feedback control is introduced to improve ride quality while satisfying the control saturation constraint introduced by limited actuator power. Sliding mode based controllers and observers are proposed in \cite{deshpande2014disturbance,liu2020adaptive} to provide robust performance for the 
active suspension system given parameter uncertainties and external disturbances. In \cite{khodadadi2018self}, the fuzzy logic self-tuning PID scheme is proposed to minimize the working space of the suspension, and improve the driver's comfort and handling stability of the vehicle suspension system. In\cite{wang2012hierarchical}, the hierarchical Takagi-Sugeno (T-S) fuzzy-neural control with structure learning is proposed for a vehicle active suspension system affected by road irregularities, to track the optimal balance point (defined as zero) of displacement of the vehicle body and vehicle wheel. MPC for nonlinear full car active suspension equipped with an advanced camera for the preview information of oncoming road is proposed in \cite{gohrle2012active} to 
improve the ride comfort and handling characteristics by utilizing lookahead information. Recently, air suspension that replaces the steel springs between the wheels and the chassis with airbags is proposed in \cite{kim2011height,kim2011fault} to achieve a good ride height tracking property in the presence of unmodeled dynamics. 

\begin{figure}[ht]
\begin{center}
\includegraphics[width=0.5\columnwidth]{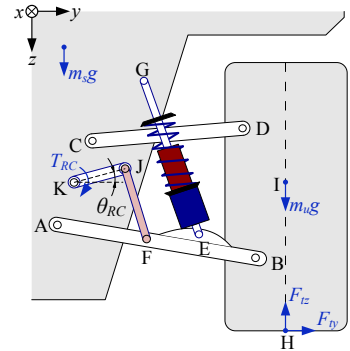}    
\caption{PALS application to a quarter car double-wishbone suspension. $m_s$ and $m_u$ are the sprung and unsprung masses, respectively. $F_{ty}$ and $F_{tz}$ are the lateral and vertical tire forces, respectively. $T_{RC}$ is the rocker torque and $\theta_{RC}$ is the rocker angle~\cite{yuchassis}.}
\label{fig1-1}
\end{center}
\end{figure}

The Parallel Active Link Suspension (PALS) is a relatively new active suspension solution \cite{yu2018parallel,yuchassis}. The schematic of a quarter-car with PALS is shown in Fig.\,\ref{fig1-1}. A rocker-pushrod assembly (`K-J-F') 
is introduced between the sprung ($m_s$) and unsprung ($m_u$) masses, in parallel with the conventional spring-damper. An active component, the rocker (`K-J'), which is driven by a rotary permanent magnet synchronous motor (PMSM) actuator, generates the torque ($T_{RC}$) acting from the chassis onto the lower wishbone (via the pushrod) to improve the performance of a double-wishbone suspension. The PALS is capable of: i) chassis attitude adjustment at low-frequency driving events, and ii) ride comfort and road holding improvement over high-frequency road surface. As compared to other existing active suspensions, the PALS offers advantages in terms of: i) negligible unsprung mass increment, ii) small sprung mass increment, iii) low actuation power requirements, with the PALS links geometrically optimized to maximize the rocker torque propagation to the vertical tire force increment, iv) simplified structure by the replacement of the conventional anti-roll bar, v) fail-safe characteristics and vi) utilization of mature and compact rotary electromehcanical actuation technology.

Previous work on the PALS has focused on the: i) the introduction of the newly vehicle suspension: Parallel Active Link Suspension \cite{yu2018control} 
ii) experimental validation with the PALS physically implemented in a quarter-car test rig, showing up to 40\% improvement in terms of the ride comfort\cite{yu2018parallel};
iii) 
development of mathematical models, including a linear equivalent model, a steady-state model and a nonlinear multi-body model of the PALS-retrofitted full car\cite{yuchassis}; 
and
iv) design of a chassis attitude control scheme which minimize the chassis pitch and roll angles\cite{yuchassis}; Despite this work, uncertainties (and component nonlinearities potentially represented as uncertainties) that commonly exist in a car equipped with the active suspension 
are ignored in the design and implementation of 
the previous PALS control schemes, which may lead to substantial deterioration of the suspension performance under certain circumstances. Examples of such uncertainties that are commonplace in practice include the variation of the payload (passengers and cargo), the nonlinear damping characteristic of suspension dampers with damper velocity, which is exhibited at long damper stroke operation, and the active actuator's frequency response during saturating operation.


The $\mu$-synthesis-based control scheme described in~\cite{zhou1998essentials} is a robust control method which goes beyond the more conventional $H_{\infty}$ control approach by allowing to take explicitly into account uncertainties. A preliminary use of the $\mu$-synthesis approach has been proposed in \cite{feng2020uncertainties} for a different type of suspension (Series Active Variable Geometry Suspension (SAVGS) \cite{arana2014series,yu2021series} and only for high frequency dynamics, with promising results. In the present work, an advanced $\mu$-synthesis-based control scheme is synthesized for the PALS, which is fundamentally different in architecture than the SAVGS, to maintain its performance improvement regardless of variations of system parameters associated with important uncertainties, and also to improve the suspension performance by accounting for nonlinearities as uncertainties during the control design.

This paper also extends the investigation from the quarter car setting \cite{yu2018parallel} to the full car with PALS, which is a significantly more challenging problem due to the higher order dynamics involved and the larger number of uncertainties in the system. 

The main contributions of the work are therefore summarized as follows: i) the characterization of important realistic uncertainties in the operation of a PALS-retrofitted full car for the purpose of designing effective active suspension control for operation under variable vehicle payload, as elaborated in the subsequent contributions, ii) the development of a $\mu$-synthesis-based control scheme for the PALS-retrofitted full car that improves its high frequency dynamics (ride comfort and road holding) in the presence of uncertainties, iii) the integration of the proposed $\mu$-synthesis control scheme with a PID scheme addressing the control of low frequency chassis attitude motions, to achieve effective multi-objective blended control for general (low and high frequency) motions under variable payload, and iv) 
numerical simulations with the adapted nonlinear multi-body model of the PALS-retrofitted full car to assess the effectiveness and efficiency of the proposed control scheme, as compared to the conventional passive suspension and the actively controlled PALS by conventional robust control that does not explicitly account for uncertainties,  while the vehicle undergoes different ISO-defined road events and variable payload.

The rest of the paper is organized as follows: 
Section 2 characterizes significant practical uncertainties of the vehicle with PALS. Section 3 designs a $\mu$-synthesis control scheme accounting for the investigated uncertainties denoted as `PALS-$\mu$' and a multi-objective control scheme with the contributions of $\mu$-synthesis and PID controls for the PALS denoted as `PALS-PID-$\mu$'. Section 4 performs numerical simulations to assess the performance of the proposed control schemes by comparing: a) the proposed `PALS-$\mu$' and a conventional scheme, an $H_{\infty}$ based vehicle vibration control scheme denoted as `PALS-$H_{\infty}$', given selected uncertainties, with respect to high frequency ride comfort and road holding improvement over the passive suspension case, and b) the proposed combined $\mu$-synthesis PID control, `PALS-PID-$\mu$', and a conventional combined $H_{\infty}$ PID control denoted as `PALS-PID-$H_{\infty}$', with respect to both the high frequency performance improvement as well as the low frequency chassis pitch and roll angle minimization. Finally, concluding remarks are discussed in Section 5.

\section{Vehicle modeling and characterization of uncertainties in PALS full car system}\label{sec:model}
In this section, a high fidelity nonlinear multi-body full car model that is developed in~\cite{yuchassis}, for both passive and active (PALS) suspension configurations, and that is used in the present work for nonlinear simulations and evaluation of 
the proposed control techniques, is briefly summarized first. Following that, a linear equivalent model of the PALS-retrofitted full car that is used in the present work for the linear control synthesis is also summarized. Finally, the identification and characterization of the sprung mass (payload) variation and suspension damping uncertainty 
arising from deviations of the actual system or approximations of nonlinear characteristics by linear counterparts, are provided.

\subsection{Nonlinear multi-body full car model and parameters} \label{2-1}
The nonlinear multi-body model of the 
full car considered in the present work has been developed in ~\cite{yuchassis}. The characteristics of a 6 degree of freedom (DOF) chassis, spinning wheels, powertrain elements (internal combustion engine, transmission gearbox, propeller shaft and differential mechanism), pinion-rack steering system, breaking system, passive suspension assembly, wheel tire force and moment system are mathematically described. Furthermore, PID controllers are synthesized separately for closed-loop longitudinal control, which coordinates the gas and braking pedal positions, and lateral control that corresponds to steering column position manipulation to implement various ISO-defined driving maneuvers. In the case of the PALS-retrofitted full car, the nonlinear model of the PALS, incorporating the rocker-pushrod assembly, the PMSM actuation, gearbox and so on, is additionally integrated at each corner of the chassis. The rocker-pushrod assembly at each corner is optimized following the procedure proposed in \cite{yu2018parallel}
to maximize the influence from rocker torque to the increment of the vertical tire force.


Previous research in \cite{yu2018parallel} suggests the suitability and performance of the PALS for heavy vehicles (for example, SUVs) with less stiff springs, as compared to, for example, high performance cars. The PALS is especially advantageous in this vehicle type as compared to other kinds of vehicle suspension (for example, with actuators acting in series to the spring) due to less energy consumption and actuation torque requirements. Therefore, in the present work, representative parameter values of this vehicle category, taken from~\cite{yuchassis}, are utilized to populate the nonlinear multi-body model already mentioned. The major parameters of the SUV both with passive suspension and PALS are given in Table \ref{tab1-1} in the appendix.

\subsection{Linear equivalent model of PALS full car}

To enable linear robust control synthesis, a linear equivalent model of the PALS full car that extends the linear equivalent quarter car model \cite{yu2018parallel} and extracts the suspension geometry nonlinearities, is employed and summarized here, as shown in Fig.\,\ref{fig1-3} and with its parameters value detailed in Table \ref{tab1-1} in the appendix.
\begin{figure}[ht]
\begin{center}
\includegraphics[width=\columnwidth]{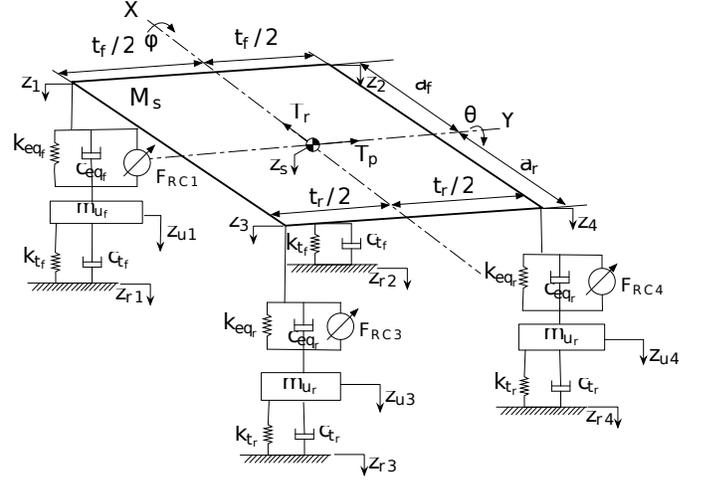}  
\vspace{-4mm}
\caption{Linear equivalent model of PALS full car. $M_s$ is the sprung mass, $T_p$ and $T_r$ are the chassis pitch and roll disturbance moments respectively, $F_{RCi}$ are the linear equivalent actuation forces at each corner, $\theta$ and $\phi$ are the pitch and roll angles respectively, and the $z$ variables denote linear vertical displacements of the points/masses indicated, while the rest of the symbols are defined in Table \ref{tab1-1} in the Appendix.}
\label{fig1-3}
\end{center}
\vspace{-6mm}
\end{figure}

The state space representation of the model can be constructed as follows:
\begin{equation}\label{1-1}
\begin{aligned}
\dot{\bar{x}} &= A\bar{x} + B\bar{u},\\
\bar{y} &= C\bar{x} + D\bar{u}.\\
\end{aligned}
\end{equation}
The state vector is chosen as follows for the minimal realisation of the linear system:
\begin{equation}\label{1-2}
\begin{aligned}
\bar{x}^T = [\dot{z}_{s},\dot{\theta},\dot{\phi}, \bm{\dot{z}_{u}}^T,\Delta \bm{l_{s}}^T,\Delta \bm{l_{t}}^T],
\end{aligned}
\end{equation}
in which: 1) $\dot{z}_{s},\dot{\theta},\dot{\phi}$ are, respectively, the linear velocity of the CMC (center of mass of the chassis) in the vertical direction, and the chassis pitch and roll rotational velocities, 2) $\bm{\dot{z}_{u}}=[\dot{z}_{u1},\dot{z}_{u2},\dot{z}_{u3},\dot{z}_{u4}]^T$ are vertical velocities of the unsprung masses ($m_{u_{(r/f)}}$) at each wheel, 3) $\Delta \bm{l_{s}}=[\Delta {l_{s1}},\Delta {l_{s2}},\Delta {l_{s3}},\Delta {l_{s4}}]^T$ are the suspension deflections at the four corners ($\Delta {l_{si}}=z_{ui}-z_{i}$, $i=1\ldots 4$), where the vertical position of the sprung mass at each wheel $z_i$ is linearised and approximated as:
\begin{equation}\label{1-3}
\begin{aligned}
z_1 = z_{s} - b_f\theta - \frac{t_f}{2}\phi, z_2 = z_{s} - b_f\theta + \frac{t_f}{2}\phi,\\
z_3 = z_{s} + b_r\theta - \frac{t_r}{2}\phi, z_4 = z_{s} + b_r\theta + \frac{t_r}{2}\phi,
\end{aligned}
\end{equation}
where $z_s$ is the linear displacement of the CMC in the vertical direction, and 4) $\Delta \bm{l_{t}} = [\Delta{l}_{t1},\Delta{l}_{t2},\Delta{l}_{t3},\Delta{l}_{t4}]^T$ are the tire deflections at each corner ($\Delta {l_{ti}}=z_{ri}-z_{ui}$, $i=1\ldots 4$). The system input vector $\bar{u}$ includes the disturbance signals and control input: 
\begin{equation}\label{1-4}
\begin{aligned}
\bar{u}^T = [\bar{w}^T,{u}^T],
\end{aligned}
\end{equation}
where the exogenous disturbance signal $\bar{w}=[{\bm{\dot{z}_{r}}}^T,T_p,T_r]^T$ consists of vertical road profiles at each wheel (where $\bm{\dot{{z}}_{r}}=[\dot{z}_{r1},\dot{z}_{r2},\dot{z}_{r3},\dot{z}_{r4}]^T$) and exogenous pitching ($T_p$) and rolling ($T_r$) torques. The control input is $u=\bm{F_{RC}}$ where $\bm{F_{RC}}=[F_{RC1},F_{RC2},F_{RC3},F_{RC4}]^T$ is the linear equivalent actuation force of the PALS at each corner. The selection of the vector of output variables, $\bar{y}$, is based on the sensors availability to achieve the control objectives as:
\begin{equation}\label{1-5}
\bar{y} = [\ddot{z}_{s},\ddot{\theta},\ddot{\phi}, \Delta \bm{l_{s}}^T,\Delta \bm{l_{t}}^T],
\end{equation} 
where the first three variables are the CMC vertical, pitch and roll accelerations, respectively.

\begin{figure}[htb!]
\begin{center}
\includegraphics[width=0.7\columnwidth]{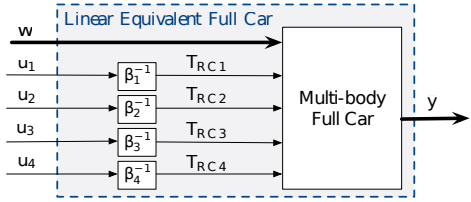}    
\caption{Conversion between linear equivalent and multi-body models of the PALS full car.}
\label{fig1-2}
\end{center}
\vspace{-6mm}
\end{figure}

In addition, as shown in Fig.\,\ref{fig1-2}, conversion functions $\beta_i$ (from each linear equivalent actuation force $F_{RCi}$ to the generated torque of the actual PALS rotary actuator) that are related to the suspension geometry variation against the suspension deflection increment, are further taken into account. That is to bridge the linear equivalent and nonlinear multi-body models of the PALS full car and compensate the varied geometry nonlinearity in terms of the suspension stroke. 

\subsection{Uncertainties in linearisation of full car PALS}

It is well known that the vehicle maneuvering can be highly uncertain, hence uncertainties are modeled to account for the discrepancy between the model used in the control design and the high-fidelity model corresponding to the real environment being investigated. The parameters of the vehicle are subject to a range of uncertainties such as external disturbances, unmodeled dynamics of rotary rocker actuator and other components, tire-road conditions, wind gusts, payload fluctuations, braking/accelerating forces, and so on.

\begin{figure}[ht]
\begin{center}
\includegraphics[width=1.0\columnwidth]{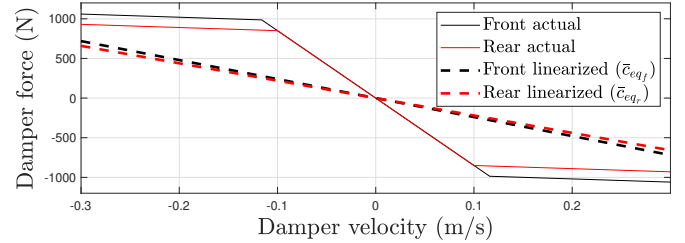}    
\vspace{-2mm}
\caption{Nonlinear (solid) and linearized (dashed) characteristics of damper force versus damper velocity for the damper units installed at the front (black) and rear (red) axles for SUV full car.}
\label{fig1-4}
\end{center}
\vspace{-4mm}
\end{figure}

For the purposes of control design of the PALS-retrofitted full car, in the present work, the most typical and significant uncertainties considered are the structured uncertainties of i) sprung mass ($M_s$), and ii) the suspension damping coefficients (${c_{eq}}_{f}$, ${c_{eq}}_{r}$) for the front and rear axle spring-damper units respectively. 
The variation of parameter $M_s$ arises from weight change in passengers and/or cargo, from the minimum (only driver), $M_{\text{min}}$, to the nominal (four male passengers), $M_{\text{nom}}$, to the maximum (seven male passengers), $M_\text{max}$, sprung mass.  
Furthermore, the 
operational speed range of the suspension dampers, which have nonlinear speed-dependent characteristics, leads to the variation of ${c_{eq}}_{f}$ and ${c_{eq}}_{r}$. 
The nonlinear damping characteristics in the real environment applied in this work are shown in Fig.\,\ref{fig1-4} (from damper manufacturer datasheets). The purpose of the linearized characteristics utilized in the linear equivalent model, as shown in Fig.\,\ref{fig1-4}, is to provide a suitable compromise between the maximum and minimum slopes of the respective nonlinear characteristics\cite{Aranaphdthesis,yu2019position}. 
The sprung mass and nonlinear dampers in the present work are therefore expressed as a nominal linear counterpart plus uncertainty. The possible ranges of values for the uncertain parameters are thus given as:
\begin{equation}\label{1-6}
\begin{aligned}
M_{s} =&~(M_{\text{nom}} + \delta M_{s})\,\text{kg},\\
{c_{eq}}_{f} =&~ (\bar{c}_{{eq}_{f}} + \delta {c_{eq}}_{f})\,\text{Ns/m},\\
{c_{eq}}_{r} =&~ (\bar{c}_{{eq}_{r}} + \delta {c_{eq}}_{r})\,\text{Ns/m},
\end{aligned}
\end{equation}
where the nominal sprung mass $M_{\text{nom}}$, and nominal suspension damping of front and rear axle $\bar{c}_{{eq}_{f}}$, $\bar{c}_{{eq}_{r}}$ are detailed in Table~\ref{tab1-1} in the Appendix. The perturbation parameters $\delta M_{s}$, $\delta {c_{eq}}_{f}$ and $\delta {c_{eq}}_{r}$ are defined as follows:
\begin{equation}\label{1-7}
\begin{aligned}
\delta M_{s} \in& 
~[-0.11M_{\text{nom}}, 0.11M_{\text{nom}}],\\
\delta {c_{eq}}_{f} \in& ~[-0.3\,\bar{c}_{{eq}_{f}}, 0.3\,\bar{c}_{{eq}_{f}}],\\
\delta {c_{eq}}_{r} \in& ~[-0.3\,\bar{c}_{{eq}_{r}}, 0.3\,\bar{c}_{{eq}_{r}}].
\end{aligned}
\end{equation}

\section{Control Methodologies Development}\label{sec:control}
This section describes the synthesis of the two control schemes for the PALS contributed by the present work. The new $\mu$-synthesis based robust control methodology (`PALS-$\mu$') is developed for high frequency ride comfort and road holding improvement. As part of the second new scheme, an existing multi-objective PID control scheme (`PALS-PID') for a PALS-retrofitted full car, proposed in \cite{yuchassis}, is utilized for the control of chassis attitude motions, particularly aiming at minimization of both pitch and roll angles. 
Finally, a frequency-dependent multi-objective control scheme (`PALS-PID-$\mu$') with the contributions of both `PALS-PID' and `PALS-$\mu$', is introduced to realize simultaneously 0-1\,Hz chassis leveling control, 1-8\,Hz vibration attenuation control and the control gains mitigation over 10\,Hz, thus to control the whole range of frequencies the vehicle operates in.

\subsection{$\mu$-synthesis-based control} \label{mu-control}
\begin{figure}[htb!]
\centering
\begin{subfigure}[]
{
\centering
\includegraphics[width=0.3\columnwidth]{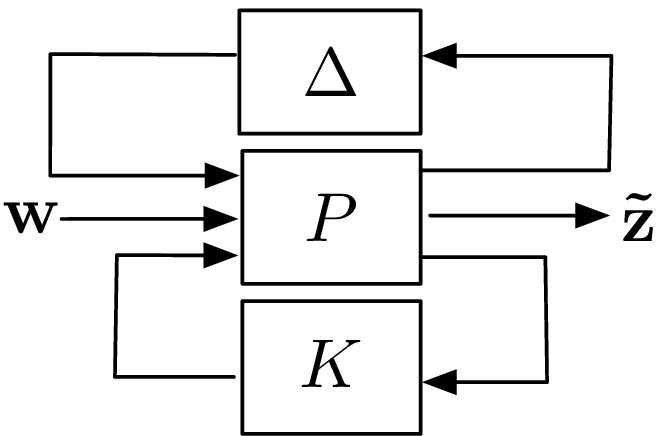}
}
\end{subfigure}\hspace{0.03\textwidth}
\begin{subfigure}[]{
\centering
\includegraphics[width=0.3\columnwidth]{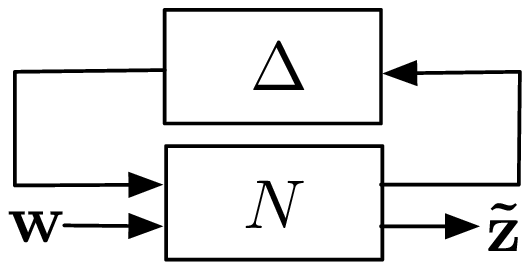}}
\end{subfigure}
\vspace{-2mm}
\caption{ $\mu$-synthesis: ($a$) extended generalized regulator, and ($b$) analysis framework \cite{zhou1998essentials}. $P$ is a time-invariant state-space representation of a linear plant, K is
a $\mu$-synthesis based controller, $\Delta$ is a structured uncertainty block diagonal matrix, ${\bf w}$ correspond to system exogenous disturbances,
and $\bar{z}$ to the cost signals to be minimized.}
\label{fig1-8}
\vspace{-4mm}
\end{figure}

The $\mu$-synthesis control allows to design a multi-variable optimal robust controller for the uncertain hand-derived linearised equivalent full car model (in Section II). It aims to attenuate the influence from external disturbances to the objective errors by synthesizing a solution with stability and performance robustness. As shown in Fig.\,\ref{fig1-8}, the $D$-$K$ iteration method is introduced to calculate a robust controller for the uncertain open loop model \cite{zhou1998essentials,chen2014mu,kaleemullah2014active}. 

The controller $K$ can be combined with the system $P$ augmented with weighting functions via a lower linear fractional transformation (LFT) $F_l$ to employ the transfer function matrix $N$. The LFT paradigm can be used in describing and analyzing the uncertain vehicle system, where $N$ is assumed to be the invariable part of the control system that contains the nominal plant and weighting values. 
As shown in Fig.\,\ref{fig1-8}, the matrix $N$ is partitioned as:
\begin{equation}
\begin{aligned}
N(s) = F_l(P(s),K(s)) = \begin{bmatrix}
N_{11}(s) & N_{12}(s)\\ 
N_{21}(s) & N_{22}(s)
\end{bmatrix},
\end{aligned}
\end{equation}

Furthermore, the upper LFT connects $\bf{w}$ and $\bf{\tilde{z}}$. Therefore, the general framework is reduced to Fig.\,\ref{fig1-8}\,($b$), and the formulation becomes:
\begin{equation}
\begin{aligned}
{\bf{\tilde{z}}} \!=\! F_u(N,\Delta)\!\cdot\!\bf{w} &\!=\!  [N_{22} \!+\! N_{21}\Delta(I\!-\!N_{11}\Delta)^{-1}\!N_{12}]\!\cdot\!\bf{w}\!,
\end{aligned}
\end{equation}
where the upper linear fractional transformation, $F_u(N,\Delta)$, is the closed-loop system from exogenous disturbance ($\bf{w}$) to cost signals ($\bf{\tilde{z}}$) to be minimized.

The structured singular value, $\mu$, is used to assess both robust stability and robust performance of the system, $N$. Mathematically,
\begin{equation}
\begin{aligned}
\mu(N)^{-1} \triangleq& \mathop {\min } \limits_{\Delta} \left \{ \bar{\sigma}(\Delta)|\det(I-N\Delta) = 0 \right \},
\end{aligned}
\end{equation}
in which $\mu$ is defined as the inverse of the maximum singular value $\bar{\sigma}$ unless no $\Delta$ makes $I-N\Delta$ singular, in which case $\mu(N) = 0$. $\mu$ provides a measure of the minimum size of the $\Delta$ block which destabilizes the $\Delta$-$N$ loop. $\Delta$ is an  uncertainty block diagonal matrix that combines the previously defined structured uncertainties, and is defined as:
\begin{equation}
\begin{aligned}
\Delta = 
\begin{bmatrix}
\delta M_{s} & 0 & 0\\ 
0 & \delta {c_{eqf}} & 0\\
0 & 0& \delta {c_{eqr}}
\end{bmatrix},
\end{aligned}
\end{equation}
where the three blocks correspond to the structured uncertainties perturbation with parameters $\delta M_s$, $\delta {c_{eqf}}$ and $\delta {c_{eqr}}$ previously defined in~\eqref{1-7}.

In $\mu$-synthesis control, the $D$-$K$ iteration method integrates two optimization problems and solves them by fixing either the variable $K(s)$ or the variable $D(s)$ by utilising $H_{\infty}$ control and $\mu$-synthesis approaches, respectively.
The upper bound of $\mu$ is given by:
\begin{equation}
\begin{aligned}
\mu(N) \leq \mathop {\min } \limits_{D \in \mathcal{D}} \bar{\sigma} (DND^{-1}),
\end{aligned}
\end{equation}
where $\mathcal{D}$ is the scaling set of nonlinear matrices ($D$) that satisfy $D\Delta=\Delta{D}$~\cite{zhou1998essentials}. The control problem is to find a controller that minimizes this aforementioned upper bound, which means solving the double minimization given by:
\begin{equation}
\begin{aligned}
\mathop {\min } \limits_{K} \left(\mathop {\min } \limits_{D\in \mathcal{D}} \left \|DN(K)D^{-1}  \right \|_{\infty}\right).
\end{aligned}
\end{equation}
The optimization problem can be solved in an iterative way using either $K(s)$ and $D(s)$, which is performed with a two-parameter minimization in a sequential way, first minimizing over $K(s)$ with $D(s)$ fixed, then minimizing pointwisely over $D(s)$ with $K(s)$ fixed. Furthermore, a minimum phase system transfer function $D(s)$ is constructed by using the magnitude of the elements of $D(j\omega)$. 
The iteration continues until $\left \|DN(K)D^{-1}  \right \|_{\infty}\leq1$ is met, or $\left \|DN(K)D^{-1}  \right \|_{\infty}$ reaches its minimum value.

\begin{figure}[htb!]
\begin{center}
\includegraphics[width=1.0
\columnwidth]{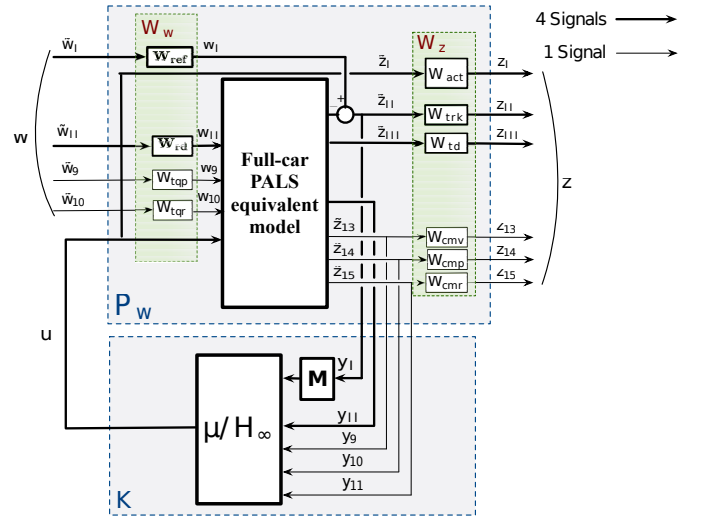}    
\vspace{-2mm}
\caption{Interconnection for $\mu$-synthesis control.}
\label{fig1-9}
\vspace{-5mm}
\end{center}
\end{figure}

Fig.\,\ref{fig1-9} shows the $\mu$-synthesis-based control configuration with the linear equivalent model of the PALS-retrofitted full car, derived in Section II ($P_{\bf w}$ is the augmented linear equivalent full car model of $P$), with disturbance signals, cost signals, manipulated variables, and measurements defined subsequently.

The unweighted normalized system disturbance inputs $\bf{w}=[{\bf{\tilde{w}_{I}}},{\bf{\tilde{w}_{II}}},\tilde{w}_{9},\tilde{w}_{10}]=[\bf{\tilde{w}_{I}},\bar{w}]$ are: 1) $\bf{\tilde{w}_{I}}=\bm{{{F}}^{(L)}_{{RC}^*}}$, the exogenous commands of the linear equivalent actuation force reference signal at low frequency level, 2) $\bf{\tilde{w}_{II}}=\bm{\dot{z}_{r}}$, the vertical road velocity of each wheel, 3) $\tilde{w}_{9}=T_p$, the equivalent pitch torque arising from load transfer due to braking/acceleration, and 4) $\tilde{w}_{10}=T_r$, the equivalent roll torque induced by load transfer when cornering. The unweighted cost signals  ${\bf \tilde{z}}=[{\bf{\tilde{z}}_{I}},{\bf{\tilde{z}}_{II}},{\bf{\tilde{z}}_{III}},{\tilde{z}}_{13},{\tilde{z}}_{14},{\tilde{z}}_{15}]$ to be minimized are: 1) $\bf{\tilde{z}}_{I}=\bm{{{F}}^{(H)}_{{RC}^*}}$, are the high frequency linear equivalent actuation forces at each corner, 2) $\bf{\tilde{z}}_{II}=\bm{{{{F}}^{(L)}_{{RC}^*}}-{F_{RC}}}$, the tracking errors of the linear equivalent actuation force, 3) $\bf{\tilde{z}}_{III}=\bm{\Delta {l}_{t}}$, the road holding related variables, which are the vertical tire deflections at each corner, and 4) the ride comfort related variables, ${\tilde{z}}_{13}$-${\tilde{z}}_{15}=[\ddot{z}_s,\ddot{\theta},\ddot{\phi}]$, which are the CMC vertical, chassis pitch and chassis roll accelerations. The measurement signals ${\bf{y}}^T = [{\bf y_{I}},{\bf y_{II}},y_9,y_{10},y_{11}]= [\bm{F_{RC}}^T,\bm{\Delta {l}_{s}}^T,\ddot{z}_s,\ddot{\theta},\ddot{\phi}]$ are selected based on the sensor availability and integration feasibility. The control signals are defined as $\bf{u} = \bm{{{F}}^{(H)}_{{RC}^*}}$.

The input weighting functions relate to the maximum expected value of the input signals ${\bf{w}}$ and ${\bf \tilde{u}}$. Constant weights chosen for disturbance signals (rate of road height, pitch and roll torque) 
aim to normalize the inputs by removing the discrepancy of physical units:

\begin{equation}
\begin{aligned}
W_{rd,\;i} = 0.25,\,\,\,\,i = 1,\dots,4,\\
W_{tqp} = 4050,\,\,W_{tqr} = 2700,
\end{aligned}
\end{equation}
where $i = 1,\dots,4$ represents the index of each vehicle corner, in the order: front left, front right, rear left and rear right. The unit influence of $\bm{F_{RC}}$ 
is normalized by introducing $W_{ref,\;i}$, which is tuned and adopted as follows:
\begin{equation}
\begin{aligned}
W_{ref,\;i} = max\left | F_{RCi} \right | = 1800,\,\,\,\,i = 1,\dots,4.
\end{aligned}
\end{equation}

$[W_{act,\;i}, W_{trk,\;i}, W_{tdi}, W_{cmv}, W_{cmp}, W_{cmr}]$ are the weighting functions working on the cost signals ${\bf \tilde{z}}$, which indicate the relative importance among different objectives. To penalise the high speeds (frequencies) for the linear equivalent actuators, $W_{act,\;i}$ are defined as high pass filters, thus ensuring the power and torque limits of the actuators are not exceeded:
\begin{equation}
\begin{aligned}
W_{act,\;i} = \frac{1}{700}\cdot\frac{\frac{1}{(2\pi\cdot10)^{2}}s^2+\frac{2}{2\pi\cdot10}s+1}{\frac{1}{(2\pi\cdot100)^{2}}s^2+\frac{2}{2\pi\cdot100}s+1},\,\,\,\,i = 1,\dots,4.
\end{aligned}
\end{equation}
$W_{trk,\;i}$ are defined approximately as integrators (with a pole at a very low frequency for better tuning). These guarantee that the desired reference signals are followed by the actuation force of the rotary actuators at low or zero frequencies, for merging with the low frequency attitude control by the PID controller, as will be described in the next Section, without overlapping with other higher frequency (2-10\,Hz) objectives including the chassis acceleration and tire deflections control:
\begin{equation}
\begin{aligned}
W_{trk,\;i}= \frac{1}{0.1}\cdot\frac{1}{\frac{1}{(2\pi\cdot0.001)}s+1},\,\,\,\,i = 1,\dots,4.
\end{aligned}
\end{equation}

Two of the main control objectives in this work are to enhance ride comfort and road holding. In terms of ride comfort, the major target is to minimize the CMC vertical, chassis pitch and chassis roll accelerations, while for road holding, the aim is to penalise the dynamic tire deflections at a certain frequency range. More specifically, according to the human body sensitivity bandwidth in the vertical direction (\,0.5-8\,Hz) and rotational direction (\,0.5-4\,Hz), respectively, detailed in\,\cite{ISO_2631-1:1997}, the cut-off frequencies of the ride comfort weighting functions, $W_{cmv}$, $W_{cmp}$, and $W_{cmr}$, are selected to be 20\,Hz, 1\,Hz, and 1\,Hz respectively. The frequency weights, $W_{tdi}$, representing road holding performance are defined (after some tuning) as band pass filters to penalise the road disturbances within 1-5\,Hz. The tuning of these weights is based on trial and error at the level of simulating with the nonlinear model after implementing the synthesized controller.

\begin{equation}\label{1-23}
\begin{aligned}
&W_{cmv} = \frac{50(\frac{1}{(2\pi\cdot160)}s+1)}{\frac{1}{(2\pi\cdot20)}s+1}\,,\,\,W_{cmp} = \frac{25}{\frac{1}{(2\pi\cdot1)}s+1},\\
&W_{cmr} = \frac{25}{\frac{1}{(2\pi\cdot1)}s+1},\\
&W_{td1} \!= \!W_{td2} \!=\! W_{td3} \!=\! W_{td4} \!=\! 
 \frac{\frac{10}{(2\pi\cdot0.001)}s\!+\!10}{\frac{1}{(2\pi)^{2}\cdot3}s^2\!+\!\frac{8}{2\pi\cdot4}\!+\!1}.
\end{aligned}
\end{equation}

In addition, block $\bf{M}$ in Fig.\,\ref{fig1-9} is introduced to include a free integrator to obtain a zero tracking error for the linear equivalent actuation force~\cite{zhou1998essentials}:
\begin{equation}
\begin{aligned}
M_{i}= \frac{\frac{1}{(2\pi\cdot20)}s+1}{s},\,\,\,\,i = 1,\dots,4.
\end{aligned}
\end{equation}

The $\mu$-synthesis-based control scheme implementation in the nonlinear multi-body model of the PALS-retrofitted full car is shown in the upper part of Fig.\,\ref{fig1-11}, where the control output actuation forces
${{{{F}}^{(H)}_{{RCi}^*}}}$ ($i=1,\ldots, 4$) defined in the linear equivalent model are converted to rocker torque references ${{{{T}}^{(H)}_{{RCi}^*}}}$ by means of functions $\beta_i$ shown in Fig.\,\ref{fig1-2}. Then the inner loop rocker torque tracking control (from $T_{{RCi}^*}$ to $T_{{RCi}}$) links the $\mu$-synthesis-based control and the mechanical system of the PALS full car, where d-q transformation and zero d-axis current control proposed in~\cite{zhou1998essentials,yuchassis} are utilized such that the three-phase PMSMs behave as DC equivalent motors, with the produced torques proportional solely to the q-axis currents.  


\subsection{Multi-objective blended control scheme} 
\begin{figure*}[htb!]
\begin{center}
\includegraphics[width=1.5\columnwidth
]{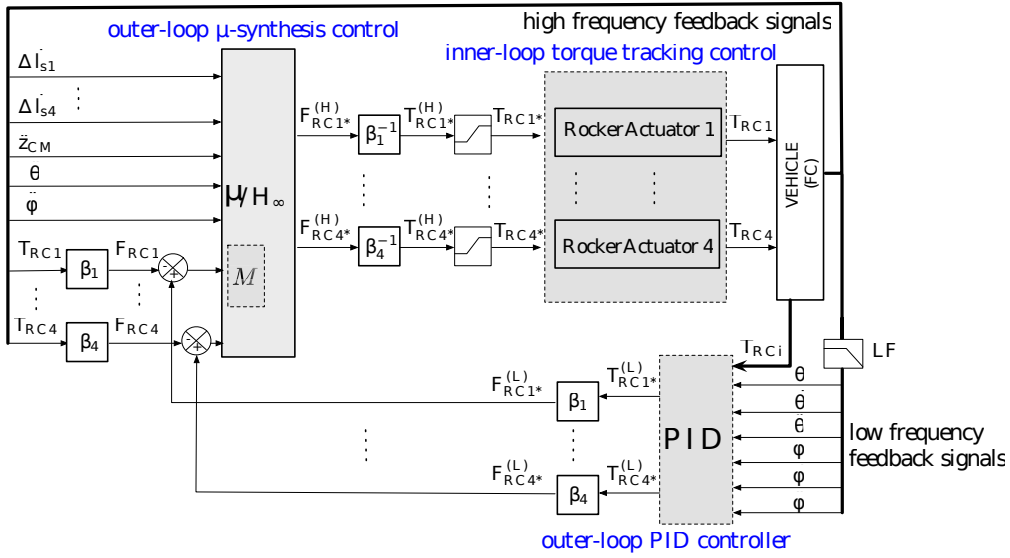}    
\vspace{-2mm}
\caption{Configuration of the multi-objective blended control scheme (`PALS-PID-$\mu$') in the PALS-retrofitted full car.}
\label{fig1-11}
\vspace{-6mm}
\end{center}
\end{figure*}

A novel multi-objective blended control scheme (`PALS-PID-$\mu$') is introduced to enable both low-frequency chassis attitude control and high frequency vibration mitigation, with the $\mu$-synthesis based control scheme (`PALS-$\mu$') combined seamlessly with the previously proposed (see \cite{yuchassis}) multi-objective PID control scheme (`PALS-PID'). The configuration of the proposed `PALS-PID-$\mu$' control scheme in the PALS-retrofitted full car is shown in the overall Fig.\,\ref{fig1-11}, which provides ride comfort and road holding enhancement during high frequency road maneuvers, given practical uncertainties, ( similarly to `PALS-$\mu$') and takes effect at the stabilization of the desired chassis attitude during low frequency maneuvers (similarly to `PALS-PID'). It is notable that the PID parameters should be retuned in `PALS-PID-$\mu$' (listed in Table~\ref{tab1-2} in the Appendix) as compared to those in `PALS-PID' due to the effect of the high frequency $\mu$-synthesis based control. Additionally, conversion functions `$\beta_i$' are also applied to bridge the linear equivalent model actuation forces (${{{{F}}^{(H)}_{RCi}}}$) and the nonlinear multi-body model rocker torques (${{{{T}}^{(H)}_{RCi}}}$).

\subsection{Benchmark control schemes}
The interconnection diagram in Fig.\,\ref{fig1-9} is also used to synthesize a conventional $H_{\infty}$ controller, `PALS-$H_{\infty}$', which does not take into account uncertainties, for benchmarking purposes. The same weighting functions as those described in Section \ref{mu-control} for the $\mu$-synthesis control are found to also be beneficial and applied to the $H_{\infty}$ control synthesis.

Moreover, the conventional $H_{\infty}$ controller is integrated with `PALS-PID' to provide the blended control `PALS-PID-$H_{\infty}$', for further benchmarking purposes with the tuning weights listed in Table~\ref{tab1-2} in the Appendix.
To perform a comprehensive comparison to the state-of-the-art of `PALS-$H_{\infty}$' and `PALS-PID-$H_{\infty}$', the $\mu$-synthesis-based controllers `PALS-$\mu$' and `PALS-PID-$\mu$' are defined with the structured uncertainties of sprung mass and suspension damping taken into account and synthesized by means of the MATLAB command \texttt{dksyn}. 


\section{Numerical Simulations and analysis}
In this section, with the nonlinear multi-body model described in Section \ref{2-1} and 
control strategies proposed in section \ref{sec:control}, a group of ISO driving maneuvers, containing, i) random road class C, ii) speed bump, iii) steady-state cornering, iv) step steer and v) brake in a turn, are tested to evaluate the PALS potential in terms of suspension performance and robustness of the synthesized controllers.

Among the above driving maneuvers, the last three are investigation cases for low frequency chassis leveling, which give rise to the control strategies of `PALS-PID', `PALS-PID-$H_{\infty}$' and `PALS-PID-$\mu$'. The other two maneuvers are utilized for the investigation of high frequency vehicle vibration control in terms of ride comfort and road holding improvement, hence two sets of control strategies, a) `PALS-$H_{\infty}$' and `PALS-$\mu$', and b) `PALS-PID-$H_{\infty}$' and `PALS-PID-$\mu$', are employed separately.

The numerical simulations are therefore performed to evaluate two aspects of performance comparison: the benefits of accounting for the suspension damping nonlinearity and sprung mass variation as parametric uncertainties in the linear control synthesis, with high frequency case studies comparison between `PALS-$H_{\infty}$' and `PALS-$\mu$' (aspect A); the benefits of combining low frequency signal tracking and high frequency vibration attenuation, given selected uncertainties, with both low frequency and high frequency investigation studies comparison between `PALS-PID-$H_{\infty}$' and `PALS-PID-$\mu$' (aspect B).

\subsection{Evaluation of aspect A}
\subsubsection{Case Study on random road}
The ISO random roads are used to simulate road unevenness. The measured vertical surface data of different road profiles, such as streets, highways as well as off-road terrain are usually described 
in terms of their power spectral density (PSD), which is defined in the frequency domain as follows~\cite{ISO_8608-2016}:
\begin{equation}
\begin{aligned}
G_{d}(n)=10^{-6}\cdot2^{2k} (\frac{n}{n_0})^{-\bar{\omega}},
\end{aligned}
\end{equation}
where $n$ is the spatial frequency and $k=2$ to $9$ corresponds to the road roughness classes $A$ to $H$ respectively. $n_0=0.1$\,cycles/m is the reference spatial frequency and $\bar{\omega}=2$ is typically a constant. In the numerical simulation environment, the PALS-retrofitted full car is driven with a forward speed of $100$\,km/h, different road profiles (of same unevenness) against traveled distance are generated for the left and right side wheels respectively, and the traveled distance delay due to the wheelbase ($\alpha_f \mbox{+} \alpha_r$) occurs in the road profile seen by the rear axle as compared to the road profile seen by the front one. In the present work, Class C road, which corresponds to a poor quality road surface, is selected for simulation to validate the improvement of the ride comfort and road holding.
\begin{figure*}
\centering
\begin{subfigure}
  \centering
  \includegraphics[width=.47\linewidth]{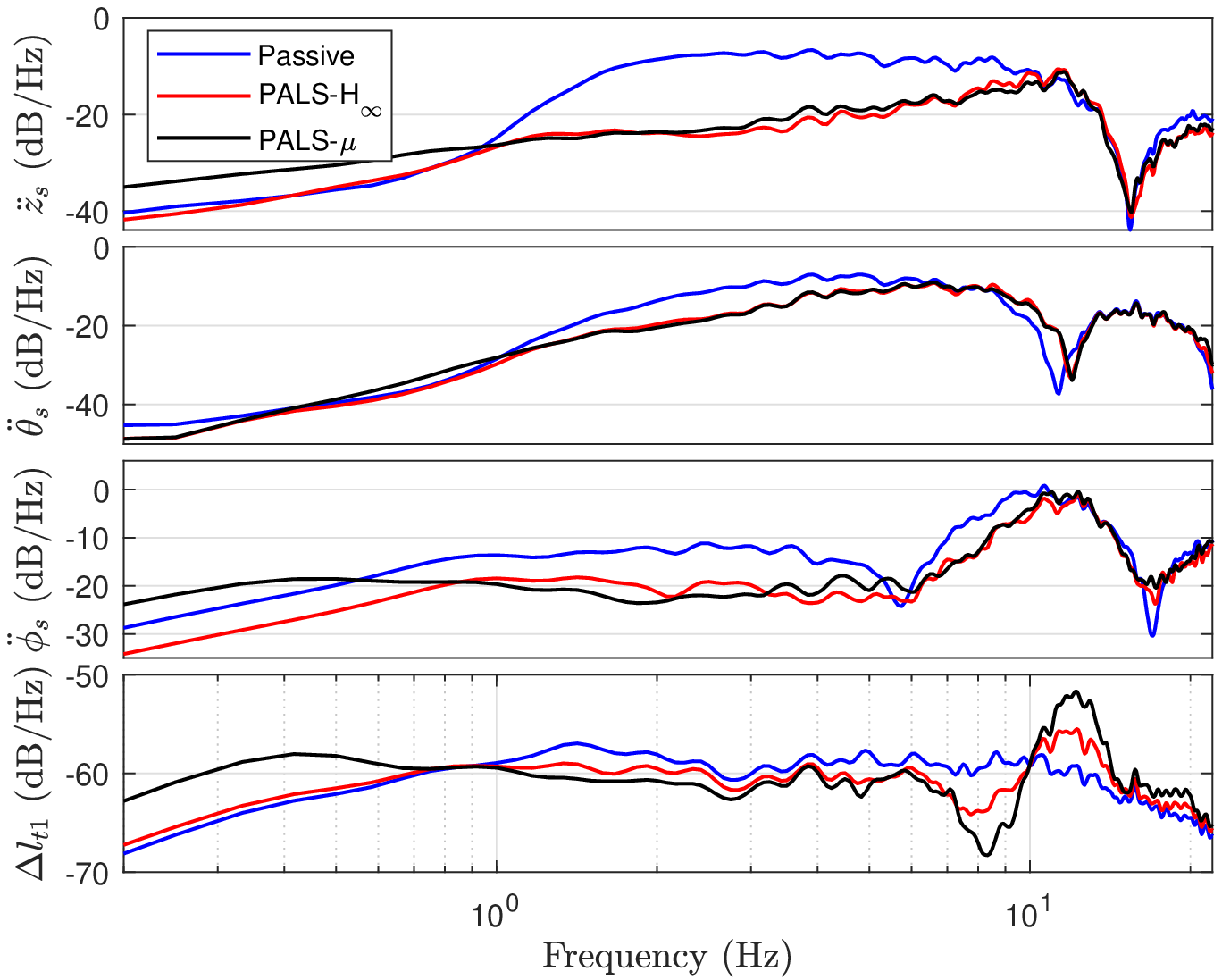}  
\end{subfigure}
\begin{subfigure}
  \centering
  \includegraphics[width=.47\linewidth]{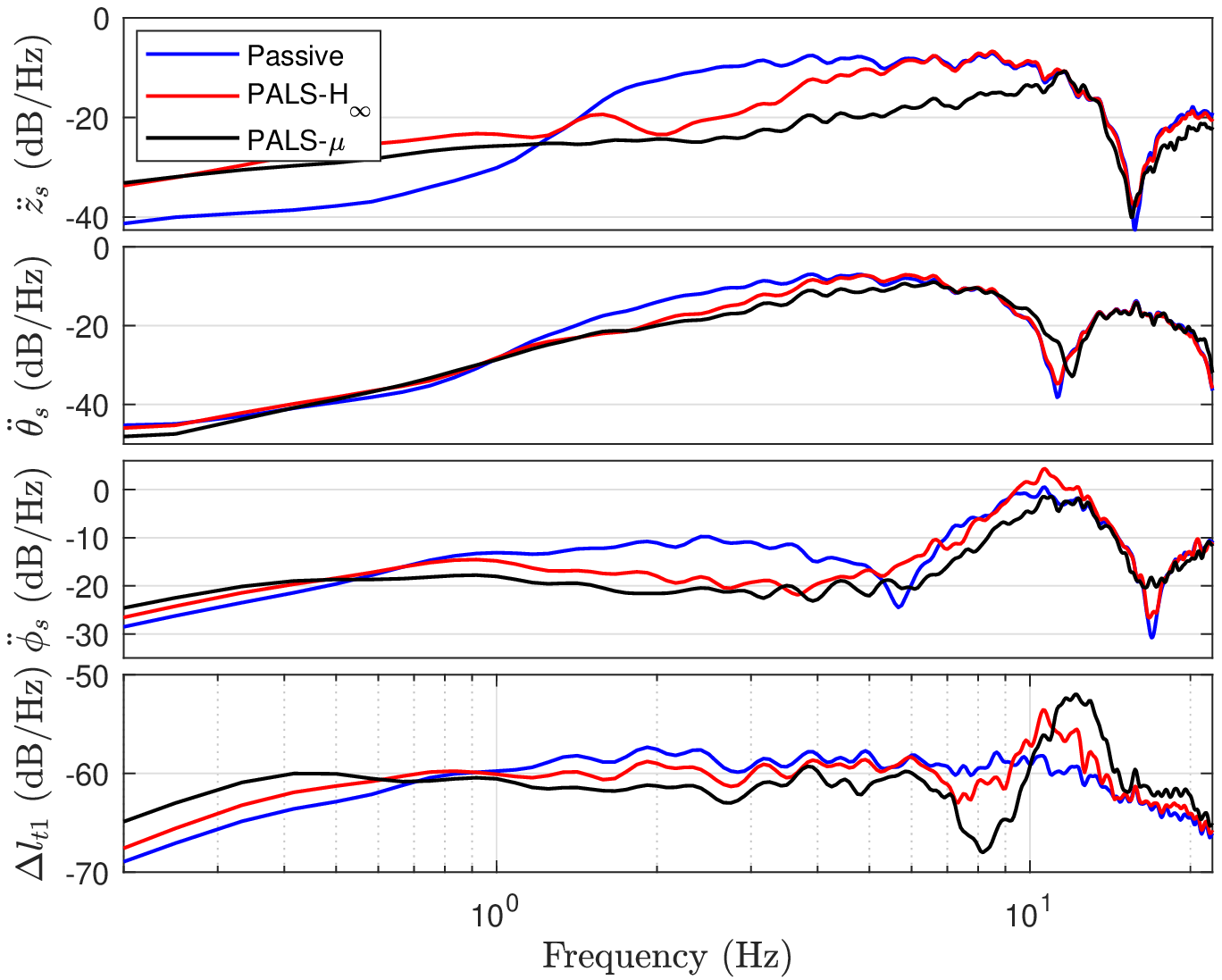}  
\end{subfigure}
\begin{subfigure}
  \centering
  \includegraphics[width=.47\linewidth]{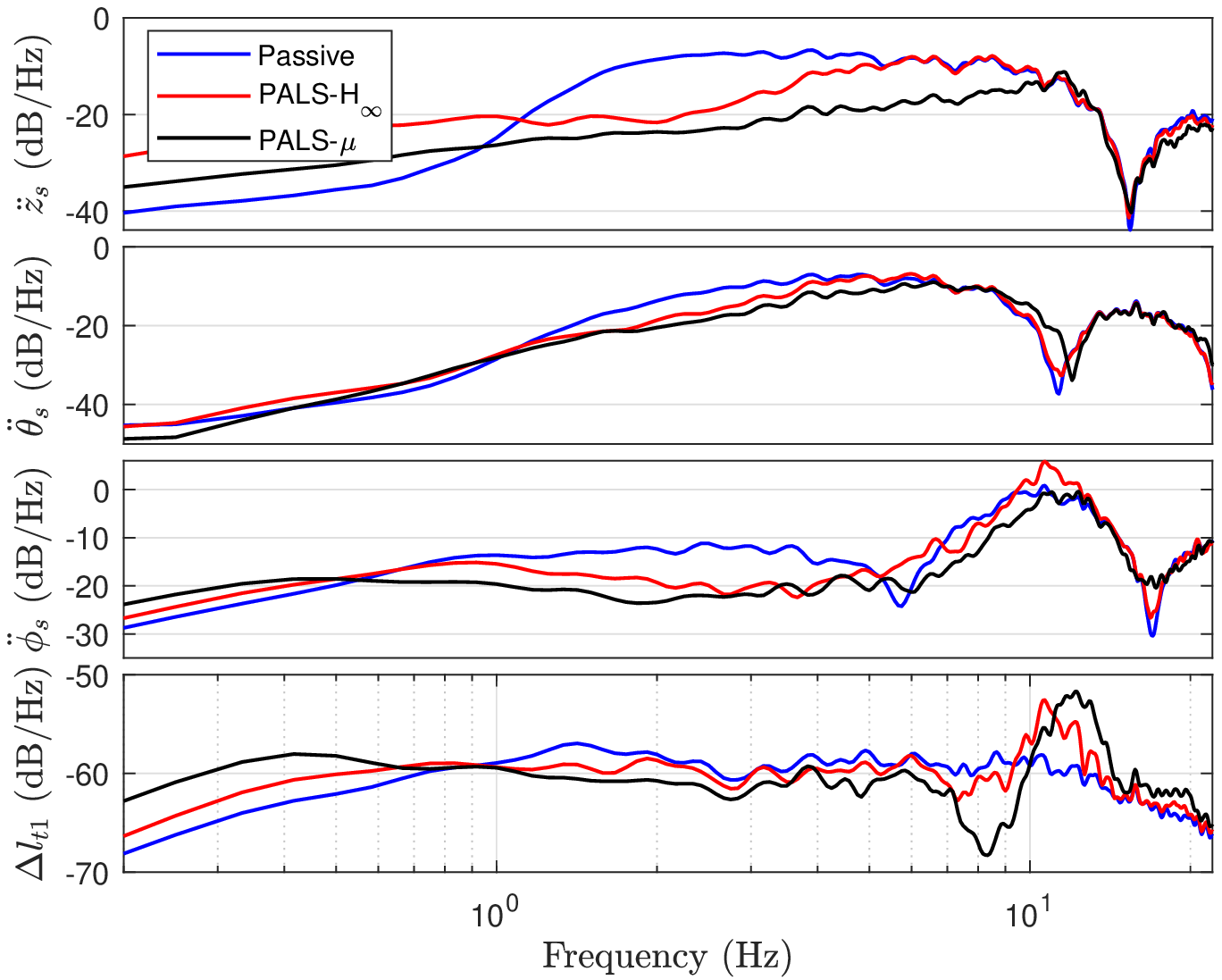}  
\end{subfigure}
\begin{subfigure}
  \centering
  \includegraphics[width=.47\linewidth]{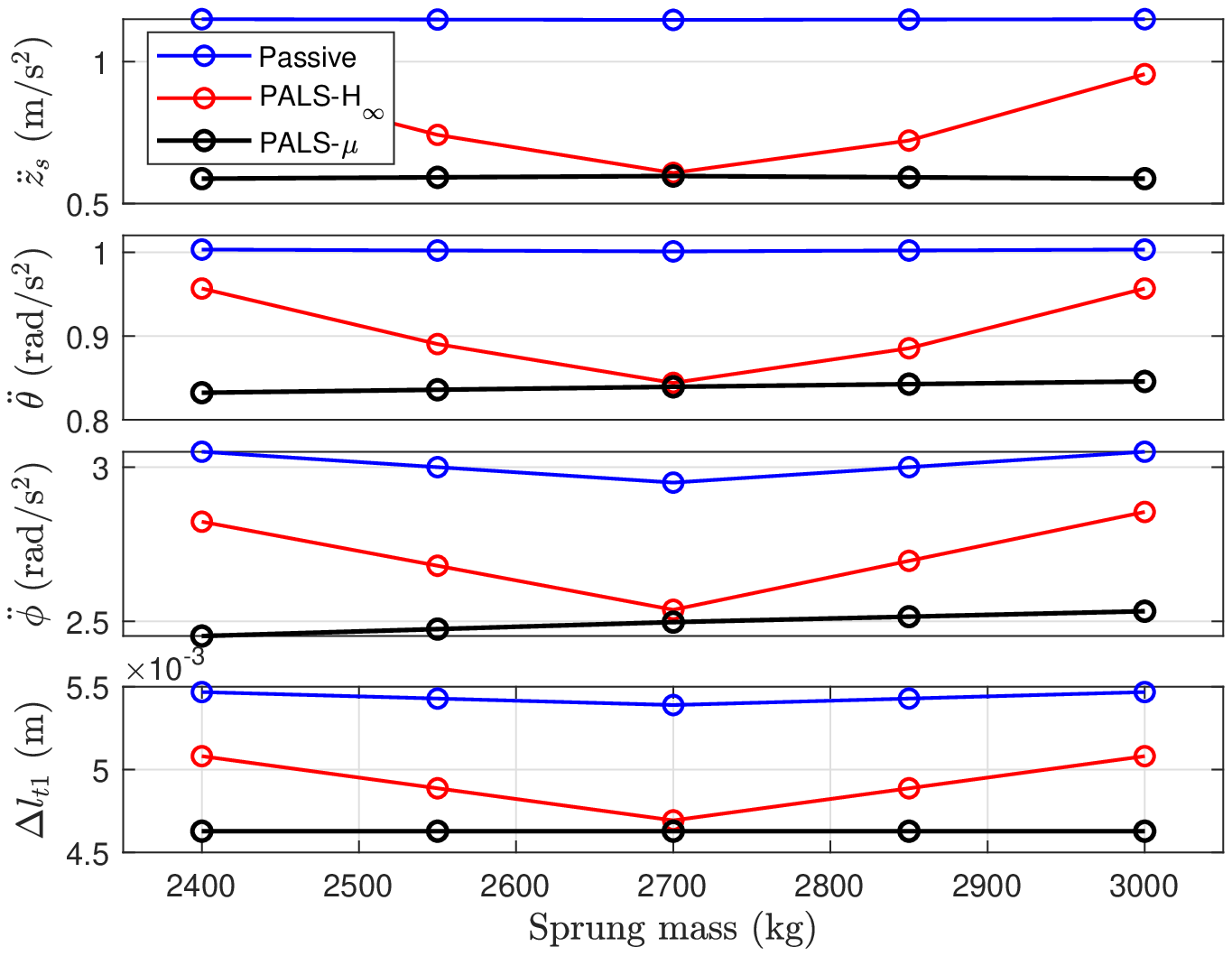}  
\end{subfigure}
\vspace{-2mm}
\caption{Numerical simulation results: 
the PSD gain of CMC vertical acceleration ($\ddot{z}_{s}$), pitch acceleration ($\ddot{\theta}_{s}$), roll acceleration ($\ddot{\phi}_{s}$) and tire deflection ($\Delta l_{t1}$) with the passive, `PALS-$H_{\infty}$' and `PALS-$\mu$' controllers in the sprung mass, $M_s$, cases of $M_{\text{nom}}$ (top-left), $M_{\text{min}}$ (top-right), and $M_{\text{max}}$ (bottom-left), and the RMS value variation of $\ddot{z}_{s}$, $\ddot{\theta}_{s}$, $\ddot{\phi}_{s}$ and $\Delta l_{t1}$ with the passive, `PALS-$H_{\infty}$' and `PALS-$\mu$' controllers for $M_s$ swept from $M_{\text{min}}$ to $M_{\text{max}}$ in 150kg increments (bottom-right) when the vehicle is running over ISO random road class C at a constant forward speed of 100\,km/h}
\label{fig1-12} 
\vspace{-6mm}
\end{figure*}

Figure.\,\ref{fig1-12} shows numerical simulation results with three different sprung mass cases of a) nominal sprung mass ($M_s\!=\!M_{\text{nom}}$), b) maximum sprung mass ($M_s\!=\!M_{\text{max}}$), and c) minimum sprung mass ($M_s\!=\!M_{\text{min}}$), for ISO random road Class C.  The evaluation indexes are selected as ride comfort related variables of the CMC vertical acceleration ($\ddot{z}_{s}$), the chassis pitch ($\ddot{\theta}_{s}$) and roll ($\ddot{\phi}_{s}$) accelerations, and the road holding related variables of the tire deflections ($\Delta l_{t1}$) at the human sensitive frequencies (1-8 Hz).

Fig.\,\ref{fig1-12} top-left reveals the PSDs of $\ddot{z}_{s}$, $\ddot{\theta}_{s}$ and $\ddot{\phi}_{s}$, and $\Delta l_{t1}$ with the PALS-retrofitted full car in the case of $M_s=M_{\text{nom}}$. As it can be seen, all the active controllers give a notably improved performance in terms of the ride comfort and the road holding as compared to the passive system, in which the $\mu$-synthesis controller `PALS-$\mu$' performs nearly the same as the `PALS-$H_{\infty}$' controller (for example, 12.4\,dB and 14.1\,dB reduction respectively at 4\,Hz in the case of CMC vertical acceleration). The reason for the marginally better performance of the `PALS-$H_{\infty}$' controller in the present case is that it is designed based on nominal conditions, as compared to the `PALS-$\mu$' controller that is designed to be more conservative to address a wider range of conditions, as will be shown next. 

The PSD plots of $\ddot{z}_{s}$, $\ddot{\theta}_{s}$ and $\ddot{\phi}_{s}$, and $\Delta l_{t1}$ in the cases of $M_s\!=\!M_{\text{min}}$ and $M_s\!=\!M_{\text{max}}$ are respectively shown in Fig.\,\ref{fig1-12} top-right and Fig.\,\ref{fig1-12} bottom-left. It can be seen that with the variation of the sprung mass, the performance enhancement of the `PALS-$H_{\infty}$' controller as compared to the passive case (for example, 4.8\,dB and 4.5\,dB reduction in terms of $\ddot{z}_{s}$ at 4\,Hz in the cases of $M_s\!=\!M_{\text{min}}$ and $M_s\!=\!M_{\text{max}}$, respectively) is not as great as when $M_s\!=\!M_{\text{nom}}$ due to the lack of robustness of the `PALS-$H_{\infty}$' controller in these case. In contrast, both sprung mass and suspension damping uncertainties take effect in these cases, in which the damper is forced to operate in a wider range of damper speeds, whereby the damper experiences more nonlinear behavior. 

Therefore, as compared to `PALS-$H_{\infty}$' the reductions of $\ddot{z}_{s}$, $\ddot{\theta}_{s}$ and $\ddot{\phi}_{s}$, and $\Delta l_{t1}$ for the `PALS-$\mu$' controller, which offers robustness against the uncertainties, can be observed (for example, -12.3\,dB and -12.4\,dB at 4\,Hz for $\ddot{z}_{s}$ as compared to the passive case for the minimum and maximum sprung mass cases, respectively). 

\begin{figure}[htb!]
\begin{center}
\includegraphics[width=8.4cm]{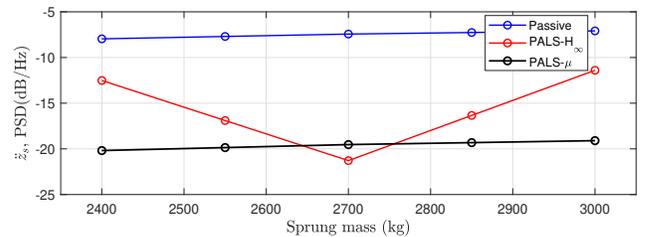}   
\vspace{-2mm}
\caption{Numerical simulation results: the PSD gain of CMC vertical acceleration ($\ddot{z}_{s}$),  at the frequency of 4\,Hz for different values of $M_s$ (swept from $M_{\text{min}}$ to $M_{\text{max}}$ in 150\,kg increments) when the vehicle is driven over an ISO random road Class C at 100\,km/h, for different cases of suspension control.}
\label{fig1-16} 
\end{center}
\vspace{-10mm}
\end{figure}

Similar conclusions can be drawn from the root mean square (RMS) quantities in the time domain with varied sprung mass, as detailed in Fig.\,\ref{fig1-12} bottom-right. It can be seen that $\ddot{z}_{s}$,$\ddot{\theta}_{s}$,$\ddot{\phi}_{s}$ and $\Delta l_{t1}$ with `PALS-$\mu$' provide the largest performance enhancement when the sprung mass is varied to its upper and lower limit, while the improvement is marginal in the nominal mass case as compared to `PALS-$H_{\infty}$'. In particular, although the PSD plots in Fig.\,\ref{fig1-12} show that $\Delta l_{t1}$ deteriorates at about 10 Hz with `PALS-$\mu$', the RMS value demonstrates that overall there is improvement. Moreover, Fig.\,\ref{fig1-16} shows the PSD value of the $\ddot{z}_{s}$ at the fixed frequency of 4\,Hz for $M_s$ swept from $M_{\text{min}}$ to $M_{\text{max}}$ in steps of 150\,kg, which demonstrates the improved robustness of the $\mu$-synthesis control scheme as compared to the $H_{\infty}$ scheme for variations in the sprung mass.

\begin{figure*}[htb!]
\centering
\begin{subfigure}
  \centering
  \includegraphics[width=.45\linewidth]{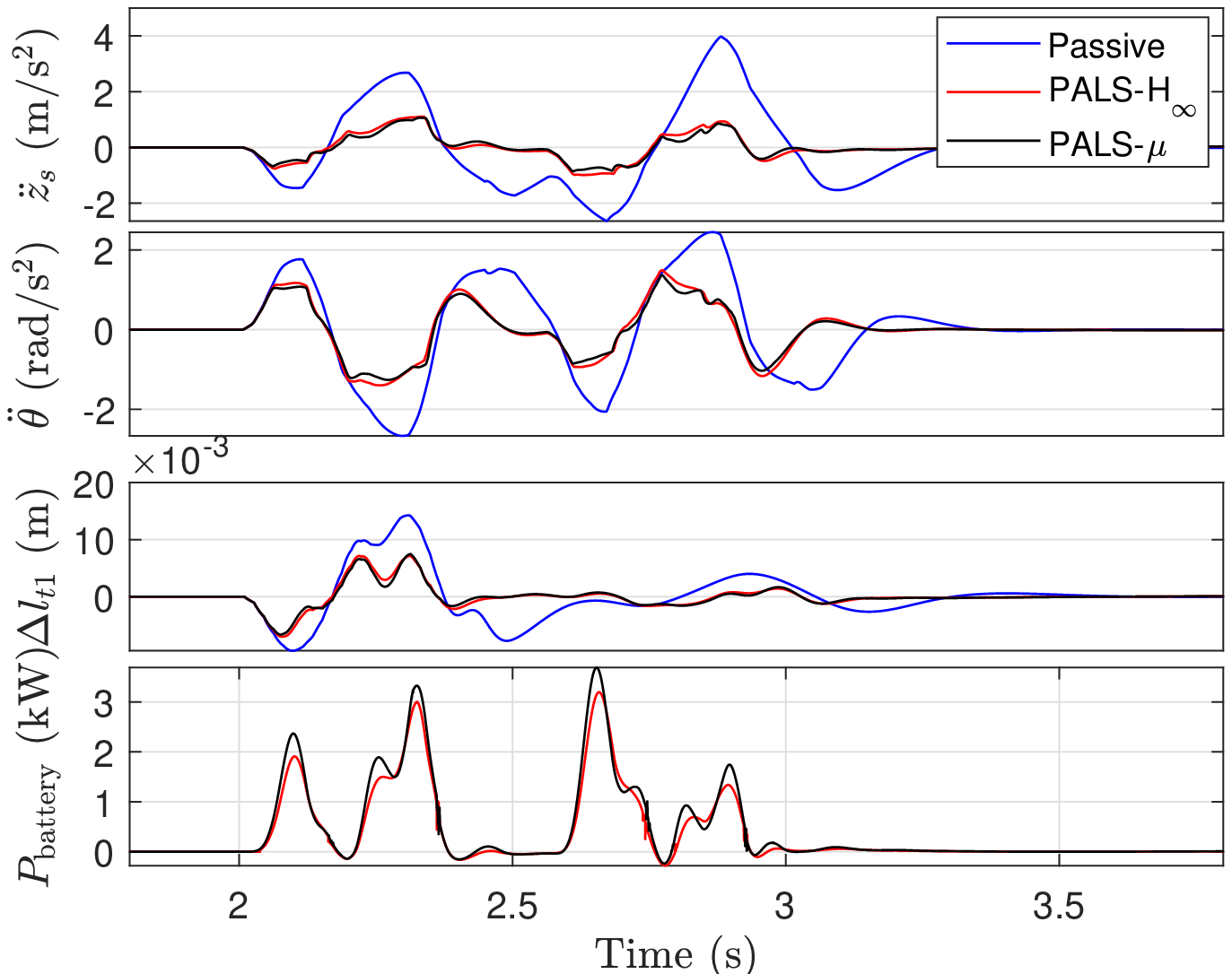}  
\end{subfigure}
\hspace{3em}
\begin{subfigure}
  \centering
  \includegraphics[width=.45\linewidth]{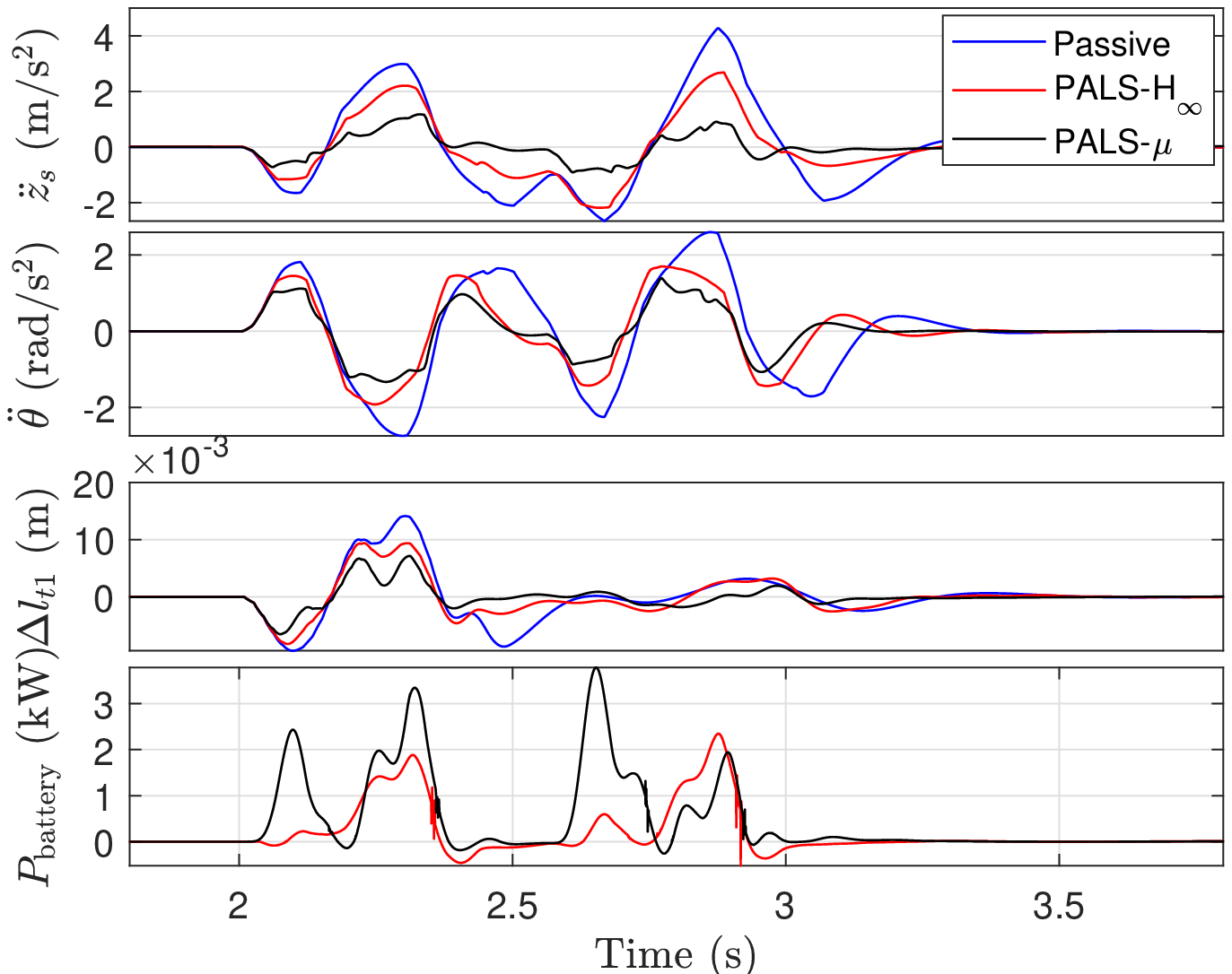}  
\end{subfigure}
\begin{subfigure}
  \centering
  \includegraphics[width=.45\linewidth]{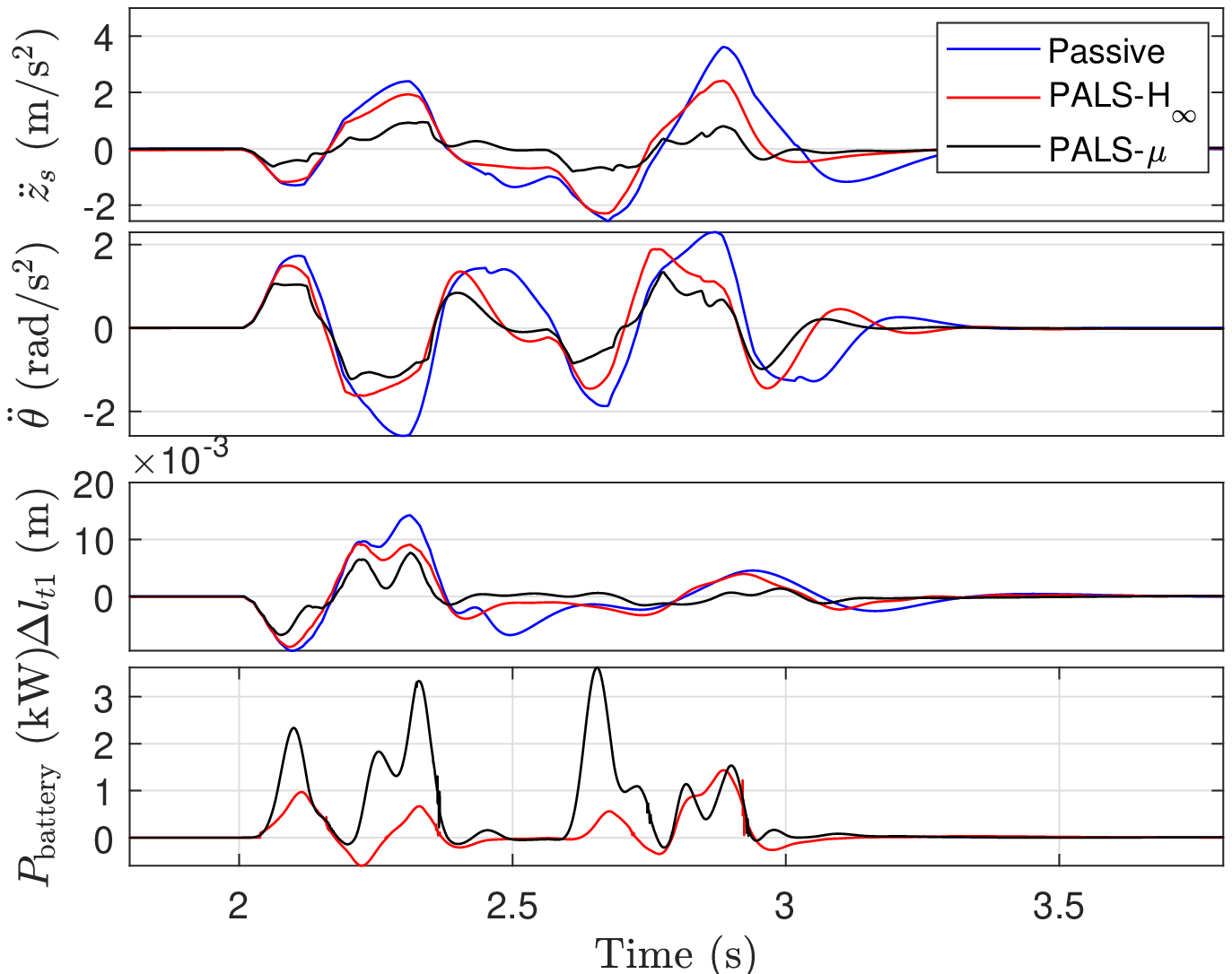}  
\end{subfigure}
\hspace{3em}
\begin{subfigure}
  \centering
  \includegraphics[width=.45\linewidth]{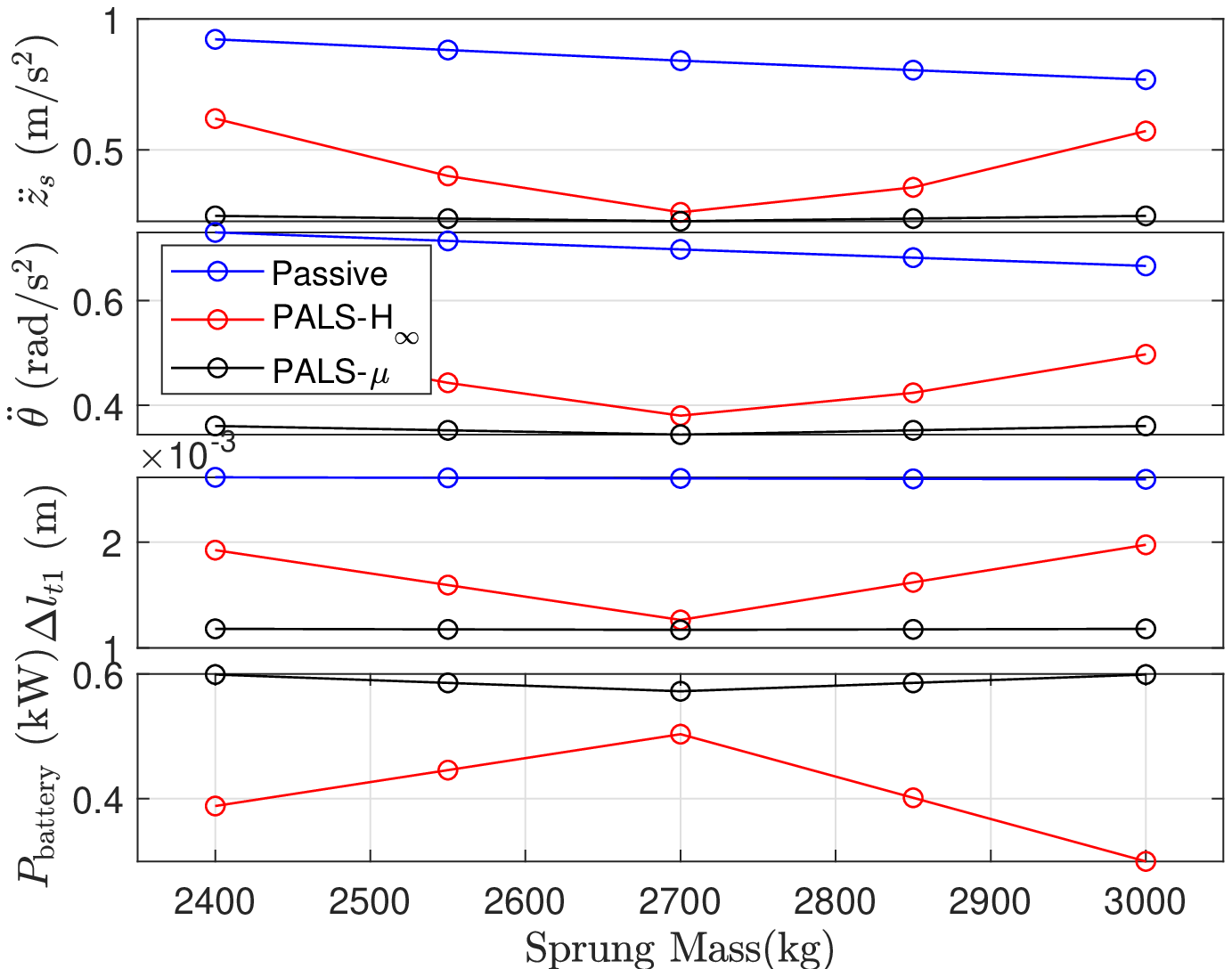}  
\end{subfigure}
\vspace{-2mm}
\caption{Numerical simulation results: the variation of CMC vertical acceleration ($\ddot{z}_{s}$), pitch acceleration ($\ddot{\theta}_{s}$), tire deflection ($\Delta l_{t1}$) and power consumption in the DC batteries ($P_{battery}$) with the passive, `PALS-$H_{\infty}$' and `PALS-$\mu$' controllers in the sprung mass, $M_s$, cases of $M_{\text{nom}}$ (top-left), $M_{\text{min}}$ (top-right), and $M_{\text{max}}$ (bottom-left), and the RMS value variation of $\ddot{z}_{s}$, $\ddot{\theta}_{s}$, $\Delta l_{t1}$ and $P_{battery}$ with the passive, `PALS-$H_{\infty}$' and `PALS-$\mu$' controllers for $M_s$ swept from $M_{\text{min}}$ to $M_{\text{max}}$ in 150kg increments (bottom-right) when the vehicle is running over a bump of 0.05\,m height and 2\,m length at a constant forward speed of 20\,km/h.
}
\label{fig1-17} 
\vspace{-6mm}
\end{figure*}

\subsubsection{Case Study on speed bump}
Speed bumps or humps are common in some roadways and are normally approximated as a raised cosine 
shape. Similarly to the random road case, the rear axle experiences a bump after a travel distance delay due to the wheel base ($l_w$) with respect to the front wheel. The mathematical representations of the front and rear wheels road heights running over a standard laterally uniform bump (0.05\,m height and 2\,m length) are presented below:
\begin{equation}\label{1-20}
\begin{aligned}
h_f &=~ 0.025(1-cos(2\pi x/2)),\\
h_r &=~ 0.025(1-cos((2\pi x-l_w)/2)),
\end{aligned}
\end{equation}
where $h_f$ and $h_r$ are functions of the travel distance $x$. In the numerical simulation environment, the PALS-retrofitted full car is driven over such a bump with a constant forward speed.

The ride comfort is mainly revealed by the CMC vertical and chassis pitch acceleration, as the roll motion is barely affected due to simultaneous excitation occurring on the two wheels of the same axle in this type of profile. The parameters of significance on road holding are dynamic tire deflections at each corner.

Numerical simulation results with the PALS-retrofitted full car at a driving speed of 20\,km/h with different cases of sprung mass are shown in Fig.\,\ref{fig1-17}. Similarly to the random road cases, the `PALS-$H_{\infty}$' controller achieves its best improvement in the case of $M_s\!\!=\!\!M_{\text{nom}}$, which contributes 70\% and 41\% reduction in the peak-to-peak (PTP) value of $\ddot{z}_{s}$ and $\Delta l_{t1}$, respectively, as compared to the passive suspension, and becomes less performing as $M_s$ is varied. The PTP values of $\ddot{z}_{s}$ and $\Delta l_{t1}$ deteriorate to 28\%, 17\% and 22\%, 18\% in the cases of $M_s\!=\!M_{\text{min}}$ and $M_s\!=\!M_{\text{max}}$, respectively, as compared to passive suspension. `PALS-$\mu$' maintains the largest ride comfort and road holding performance enhancement for the whole $M_s$ range cases, which for example reduces the PTP value of $\ddot{z}_{s}$ and $\Delta l_{t1}$ by approximately 71\% and 39\% for the $M_{\text{nom}}$ case (top-left plot in Fig.\,\ref{fig1-17}), 73\% and 40\% for the $M_{\text{min}}$ case (top-right plot in Fig.\,\ref{fig1-17}), and 69\% and 38\% for the $M_{\text{max}}$ case (bottom-left plot in Fig.\,\ref{fig1-17}), which further illustrate the significant enhancement of $\mu$-synthesis controller over the $H_{\infty}$ controller. The improvement offered by PALS-$\mu$ over the passive and $H_{\infty}$ controller cases, in terms of time-domain RMS values, is further observed in Fig.\,\ref{fig1-17} bottom-right. The RMS value of total power consumption in the DC batteries for `PALS-$H_{\infty}$' is 0.39\,kW, 0.51\,kW and 0.30\,kW in the $M_{\text{min}}$, $M_{\text{nom}}$ and $M_{\text{max}}$ cases, respectively. The `PALS-$\mu$' dissipates similar power in the $M_{\text{nom}}$ case as compared to `PALS-$H_{\infty}$' and in order to achieve its better performance for the masses away from $M_{\text{nom}}$, the `PALS-$\mu$' dissipates more power with 0.58\,kW and 0.59\,kW in the $M_{\text{min}}$ and $M_{\text{max}}$ cases, respectively (referring to top-right and bottom-left plots in Fig.\,\ref{fig1-17}).


\subsection{Evaluation of aspect B}
The purpose of evaluating aspect B is to emphasize the improvement of the robust control scheme `PALS-PID-$\mu$' over the conventional control scheme `PALS-PID-$H_{\infty}$', with selected uncertainties taken into account, in terms of low frequency vehicle attitude tracking and high frequency vibration attenuation.
\subsubsection{Case Study on steady-state cornering}
The open-loop test methods defined in \cite{iso20124138} determine how the vehicle behaves in a steady-state circular driving condition. In the nonlinear simulation environment, the PALS-retrofitted full car is driven with a constant forward speed of 100\,km/h, with the steering wheel angle increased linearly from 0 to 60 degrees over a time period of 400s.

\begin{figure}[htb!]
\begin{center}
\includegraphics[width=8.4cm]{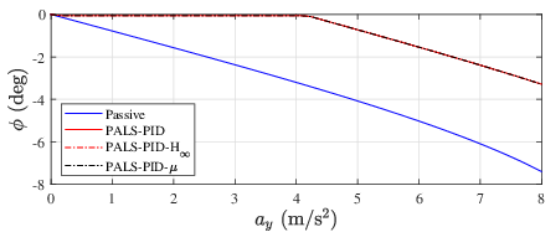}    
\vspace{-2mm}
\caption{Numerical simulation results: the chassis roll angle ($\phi$) for the $M_{\text{nom}}$ case when the vehicle is undergoing an ISO steady-state cornering at 100 km/h, for different cases of active suspension control.}
\label{fig1-20}
\vspace{-4mm}
\end{center}
\end{figure}

The variation of the roll angle $\phi$ against the lateral acceleration $a_y$ is shown in Fig.\,\ref{fig1-20}, where the `PALS-PID' achieves chassis leveling when cornering with $a_y$ up to approximately 4\,m/s$^2$. The `PALS-PID-$H_{\infty}$' and `PALS-PID-$\mu$' control strategies present the same performance enhancement over the passive suspension as compared to `PALS-PID', due to the successful tracking of the saturating reference actuation forces, ${{{F}}^{(L)}_{{RCi}^*}}$, at low frequencies coming from the PID part of both hybrid controls.

\subsubsection{Case Study on Step Steer}
\begin{figure}[htb!]
\begin{center}
\includegraphics[width=8cm]{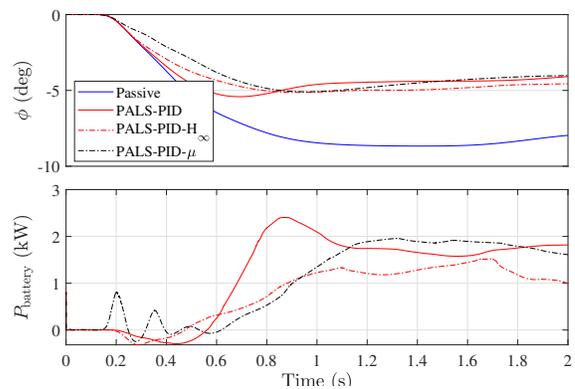}    
\vspace{-2mm}
\caption{Numerical simulation results: the chassis roll angle ($\phi$) and total power consumption in the DC batteries ($P_{battery}$) for the $M_{\text{nom}}$ case when the vehicle is undergoing an ISO step steer at 100 km/h, for different cases of active suspension control.}
\label{fig1-21}
\vspace{-6mm}
\end{center}
\end{figure}

The open-loop test methods defined in \cite{iso2011road} investigate the transient response behavior of passenger vehicles. Here, the PALS-retrofitted full car is driven at a constant forward speed of 100\,km/h, with the steering wheel angle increasing at a constant rate of 500 deg/s from 0 to 48.6 deg such that the vehicle  stabilizes at a lateral acceleration of $a_y$\,=\,8\,m/s$^2$. The variation of roll angle $\phi$ and the total power consumption in the DC batteries, $P_{battery}$, is indicated in Fig.\,\ref{fig1-21}, where the `PALS-PID' provides 42\%  mitigation of roll angle over the passive suspension, with the peak value of $P_{battery}$ being approximately 2.5\,kW. The `PALS-PID-$H_{\infty}$' produces slightly less performing results of roll angle RMS value attenuation than `PALS-PID' with a 1.5\,kW $P_{battery}$ peak value and 40\% RMS value attenuation of the roll angle as compared to the passive suspension. However, the `PALS-PID-$\mu$' presents similar roll angle RMS value reduction as compared to `PALS-PID' (\,41.87\% reduction as compared to passive case) but with lower battery power (2\,kW $P_{battery}$ peak value).

\subsubsection{Case Study on Brake in a Turn}
The open-loop test methods defined in\cite{ISO_7975-2006} determine how the steady-state circular response of a vehicle is altered by a braking action. Here, the PALS-retrofitted full car is initially driven in a circular path of 100\,m radius at a constant lateral acceleration of 5\,m/s$^2$, corresponding to a constant forward speed of 80\,km/h, then the steering wheel is fixed and brakes applied to enable the vehicle to slow down at a constant deceleration of $a_x\!=\!\mbox{-}$5\,m/s$^2$.

\begin{figure}[htb!]
\begin{center}
\includegraphics[width=8.4cm]{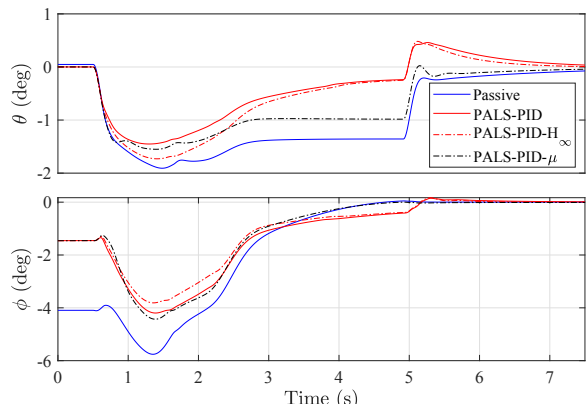}
\vspace{-2mm}
\caption{Numerical simulation results: the chassis pitch angle ($\theta$) and chassis roll angle ($\phi$) for the $M_{\text{nom}}$ case when the vehicle is undergoing an ISO brake in turn at initial longitudinal speed 80 km/h, for different cases of active suspension control.}
\label{fig1-22}
\vspace{-5mm}
\end{center}
\end{figure}

The variations of the roll angle, $\phi$, and pitch angle, $\theta$, are shown in Fig.\,\ref{fig1-22}. It can be seen that as compared to the passive suspension, the `PALS-PID' has less overshoot but it takes a longer time to stabilize in terms of roll angle performance, and its average pitch angle is reduced from \mbox{-1.5\,$\degree$} to \mbox{-0.4\,$\degree$} with a larger overshoot at approximately 5\,s before it comes to a stop. `PALS-PID-$H_{\infty}$' has a similar performance to that of `PALS-PID' for the whole time except for marginally larger pitch and smaller roll angle overshoots in the time period of 1\,s-2\,s. `PALS-PID-$\mu$' provides pitch angle response close to that of `PALS-PID' and `PALS-PID-$H_{\infty}$', which indicates the improvement over the passive suspension. In terms of roll angle performance, `PALS-PID-$\mu$' shows a similar response from 1\,s to 3.5\,s as compared to the other two active cases, while after 3.5\,s, it follows the passive suspension response ending at 5\,s which is faster than the other two active cases in that time period.

\subsubsection{Case Study on high-frequency driving maneuvers}
The two multi-objective blended control strategies `PALS-PID-$H_{\infty}$' and `PALS-PID-$\mu$' are tried while following the previously defined high frequency driving maneuvers (speed bump and random road Class C) and compared to `PALS-$H_{\infty}$' and `PALS-$\mu$' for the same maneuvers to demonstrate the effectiveness and efficiency of the proposed
`PALS-PID-$\mu$' control.

\begin{figure}[htb!]
\begin{center}
\includegraphics[width=8.4cm]{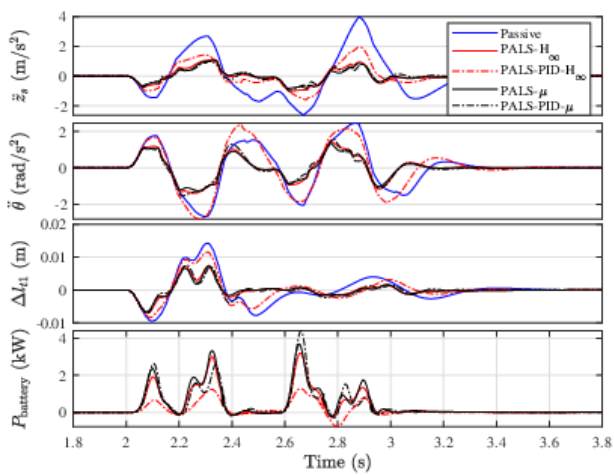}    
\vspace{-2mm}
\caption{Numerical simulation results: the time-domain value of CMC vertical acceleration ($\ddot{z}_{s}$), pitch acceleration ($\ddot{\theta}$), tire deflection $\Delta l_{t1}$ and power consumption in the DC batteries ($P_{battery}$) for the $M_{\text{nom}}$ case when the vehicle is driven over an ISO speed bump at 20 km/h, for different cases of suspension control.}
\label{fig1-23}
\vspace{-5mm}
\end{center}
\end{figure}

Numerical simulation results of the speed bump case for the nominal sprung mass case, $M_\text{nom}$, are shown in Fig.\,\ref{fig1-23}, where `PALS-$H_{\infty}$' and `PALS-$\mu$' achieve the best performance in terms of ride comfort and road holding (as seen by the vertical, pitch acceleration, and tire deflection responses) as compared to the passive case, with their performance and power consumption being very similar. Due to the compromise of the high frequency performance because of the influence of the PID control part in `PALS-PID-$H_{\infty}$', it performs worse than `PALS-$H_{\infty}$' especially for the pitch acceleration, while also consuming less power than `PALS-$H_{\infty}$'. In contrast, `PALS-PID-$\mu$' offers nearly the same significant improvement in terms of ride comfort and road holding with slightly less power consumption as compared to `PALS-$\mu$', as shown in the $P_{battery}$ plot of Fig.\,\ref{fig1-23}, which indicates that the compromise in the performance of `PALS-PID-$\mu$' at higher frequencies due to its low frequency-intended PID control aspect, is much less than the corresponding compromise for `PALS-PID-$H_{\infty}$'.

\begin{figure}[htb!]
\begin{center}
\includegraphics[width=8.4cm]{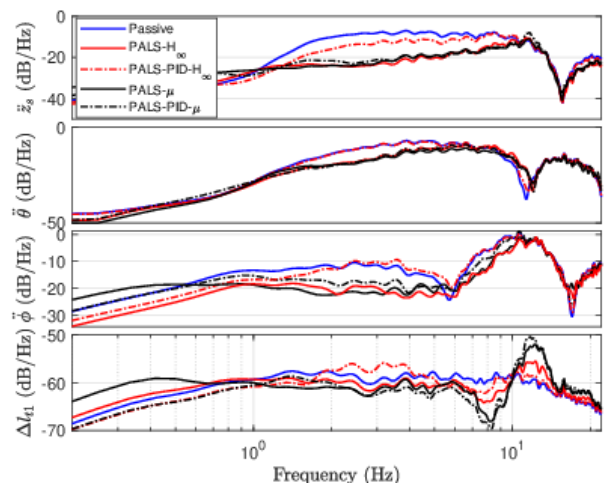}    
\vspace{-2mm}
\caption{Numerical simulation results: the PSDs of CMC vertical acceleration ($\ddot{z}_{s}$), pitch acceleration ($\ddot{\theta}$), roll acceleration ($\ddot{\phi}$) and tire deflection $\Delta l_{t1}$ for the $M_{\text{nom}}$ case when the vehicle is driven over an ISO random road Class C at 100\,km/h, for different cases of suspension control.}
\vspace{-6mm}
\label{fig1-24}
\end{center}
\end{figure}

\begin{figure}[htb!]
\begin{center}
\includegraphics[width=8.4cm]{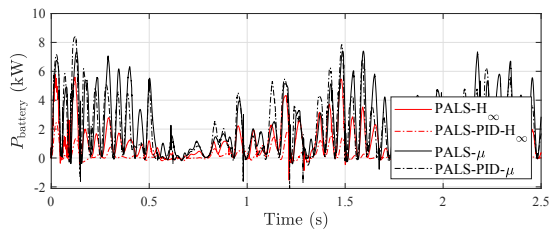}    
\vspace{-2mm}
\caption{Numerical simulation results: the total power consumption in the DC batteries ($P_{battery}$) in a 2.5\,s time history for the $M_{\text{nom}}$ case when the vehicle is driven over an ISO random road Class C at 100\,km/h, for different cases of active suspension control.}
\vspace{-6mm}
\label{fig1-25}
\end{center}
\end{figure}

Similar conclusions can be drawn from Fig.\,\ref{fig1-24} of the PSD plot in road Class C maneuver, where again `PALS-$H_{\infty}$' and `PALS-$\mu$' achieve the best performance in terms of ride comfort and road holding (reduction of PSD gains for all variables in Fig.\,\ref{fig1-24} in the frequency range of interest), with the `PALS-$\mu$' having an edge as compared to `PALS-$H_{\infty}$' in some of the metrics in Fig.\,\ref{fig1-24} and therefore being the ultimate best case. Also similarly, `PALS-PID-$H_{\infty}$' produces less ride comfort and road holding improvement than `PALS-$H_{\infty}$' as compared to the passive case. On the other hand, `PALS-PID-$\mu$' still provides similar ride comfort and road holding performance as compared to `PALS-$\mu$', which indicates the significant improvement over `PALS-PID-$H_{\infty}$'. The total power consumption in the batteries for the $M_{\text{nom}}$ case with different active suspension controllers is shown in Fig.\,\ref{fig1-25}, where it can be seen that the $\mu$-synthesis controllers consume more power than the $H_\infty$ controllers in order to provide the edge in the performance illustrated in Fig.\,\ref{fig1-24} over the $H_\infty$ controllers. It is also clear that the power consumption of `PALS-PID-$\mu$' is similar to the that of `PALS-$\mu$', further indicating their similar behavior at high frequencies.

\begin{figure}[htb!]
\begin{center}
\includegraphics[width=7.35cm]{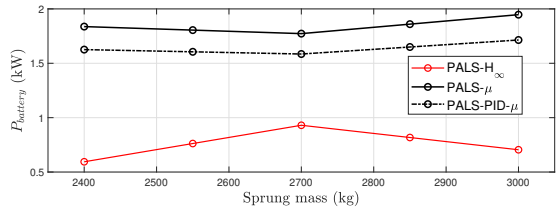}    
\vspace{-2mm}
\caption{The average value variation of total power consumption in the DC batteries when the vehicle is driven over an ISO random road Class C at 100\,km/h and $M_s$ is swept from $M_{\text{min}}$ to $M_{\text{max}}$ in 150\,kg increments, for the cases of `PALS-$H_{\infty}$',`PALS-$\mu$' and `PALS-PID-$\mu$' controllers.}
\vspace{-6mm}
\label{fig1-26}
\end{center}
\end{figure}

Fig.\,\ref{fig1-26} further illustrates the average value of $P_{battery}$ for $M_{s}$ swept from $M_{\text{min}}$ to $M_{\text{max}}$.
It is clear that `PALS-PID-$\mu$' consumes slightly less power as compared to `PALS-$\mu$', which is also consistent with providing slightly less control performance, as seen, for example, in Fig.\,\ref{fig1-24}. These two $\mu$-synthesis controllers provide in a similar manner performance improvement (see below) and power consumption increment with varied sprung mass, while the `PALS-$H_{\infty}$', which is overall somewhat less power consuming that both the $\mu$-synthesis controllers, becomes even less power consuming (and consistently less performing) as $M_{s}$ is increased or decreased away from $M_{\text{nom}}$. 

\begin{figure}[htb!]
\begin{center}
\includegraphics[width=8.4cm]{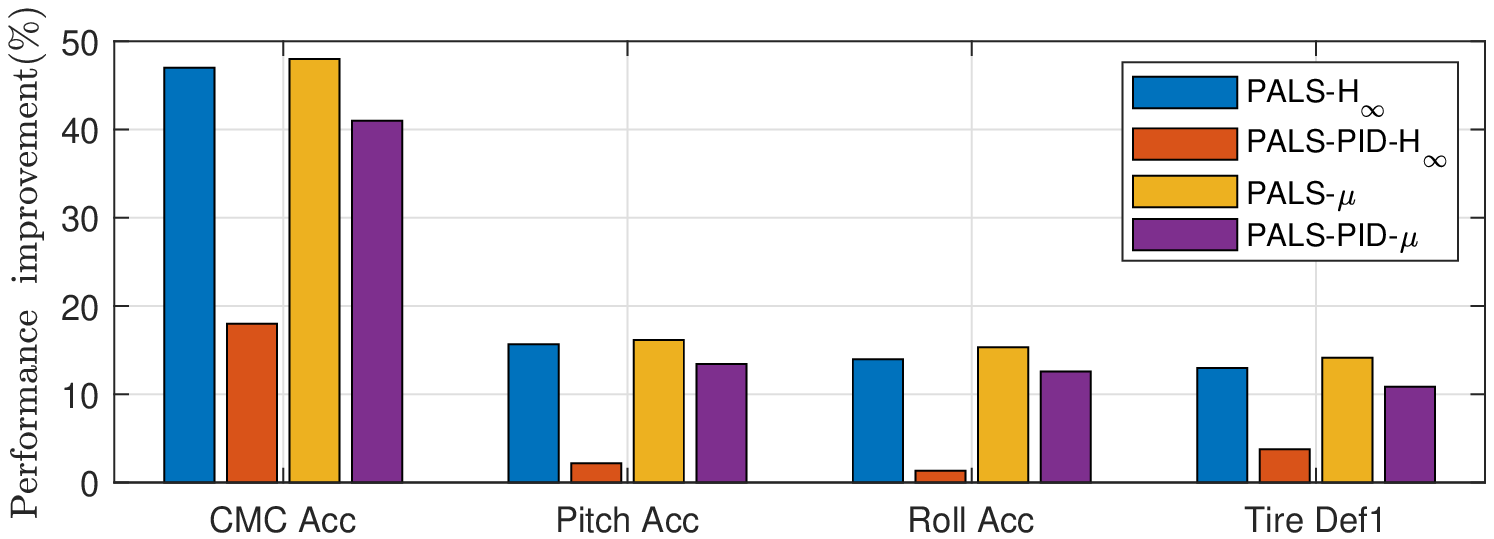}    
\includegraphics[width=8.4cm]{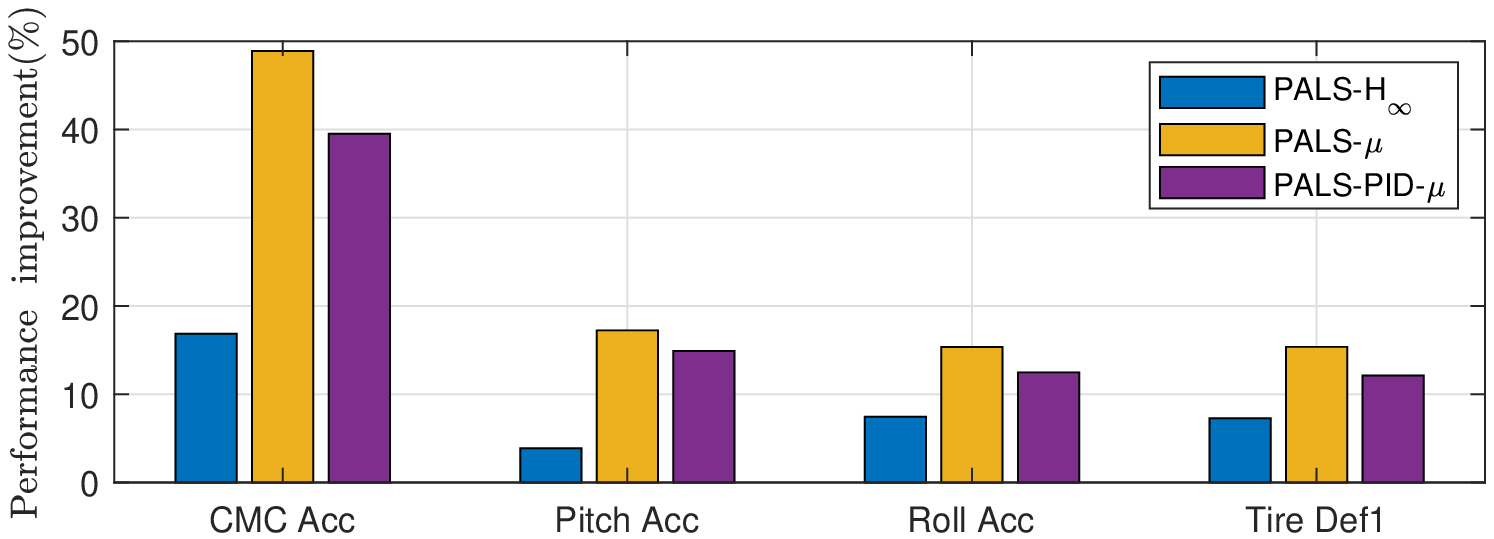}    
\includegraphics[width=8.4cm]{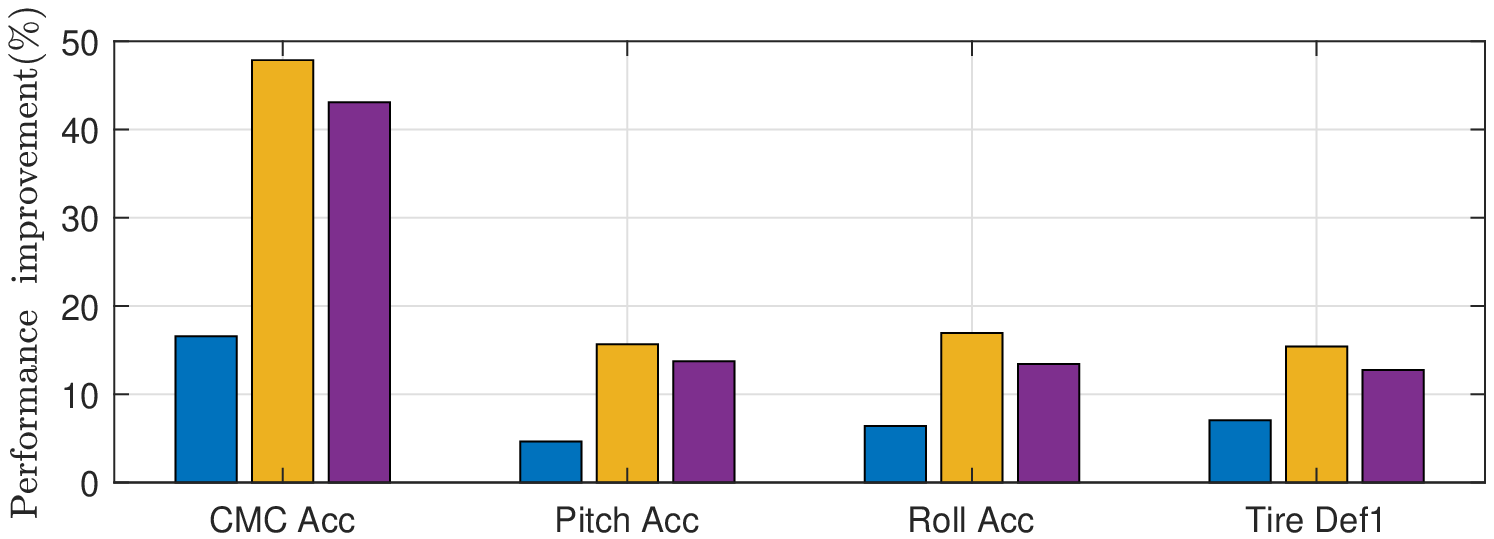}    
\vspace{-2mm}
\caption{Numerical simulation results: the RMS value performance improvement of CMC vertical acceleration ($\ddot{z}_{s}$), pitch acceleration ($\ddot{\theta}$), roll acceleration ($\ddot{\phi}$) and tire deflection ($\Delta l_{t1}$), as a percentage of the passive case performance, in the sprung mass, $M_s$, cases of $M_{\text{nom}}$ (top), $M_{\text{min}}$ (middle), and $M_{\text{max}}$ (bottom) when the vehicle is driven over an ISO random road Class C at 100\,km/h, for different cases of suspension control. The precise calculation of the height of the bars is $\frac{\text{metric}_\text{passive}-\text{metric}_\text{active}}{\text{metric}_\text{passive}}$, where `metric' is the RMS value of one of $\ddot{z}_{s}$, $\theta$, $\phi$, and $\Delta l_{t1}$, and `active' is one of the active suspension controller cases.}
\label{fig1-27}
\vspace{-6mm}
\end{center}
\end{figure}

The performance improvements alluded to above are demonstrated in Fig.\,\ref{fig1-27} in terms of body accelerations and tire deflection RMS values as percentage improvements over the passive case, for different cases of sprung mass. It can be clearly seen that even in the nominal sprung mass case `PALS-$\mu$' universally outperforms `PALS-$H_{\infty}$', marginally in most cases but significantly in roll acceleration, while in the case of the multi-objective schemes, `PALS-PID-$\mu$' performs significantly better than `PALS-PID-$H_{\infty}$' in all metrics. It is therefore clear that the performance deterioration suffered by combining the low and high frequency objectives is much less in `PALS-PID-$\mu$' with respect to `PALS-$\mu$', as compared to `PALS-PID-$H_{\infty}$' with respect to `PALS-$H_{\infty}$'. This can be attributed to the robustness of the $\mu$-synthesis approach to a more realistic class of perturbations in comparison to the $H_{\infty}$ control methodology. The benefits of $\mu$-synthesis over $H_{\infty}$ control approaches when the sprung mass is different from the nominal one are even higher as compared to the nominal sprung mass case and are clearly demonstrated in the middle and bottom plots of Fig.\,\ref{fig1-27}.

\section{Conclusions}
The recently proposed mechatronic suspension of the Parallel Active Link Suspension (PALS) is investigated in the application to a SUV full car, with sprung mass and suspension damping uncertainties taken into consideration in the suspension control design, revealing promising potential for both low frequency chassis leveling and high frequency ride comfort and road holding improvement under variable payload. 

The $\mu$-synthesis control strategy is utilized for PALS high-frequency application, with essential improvement over the passive suspension system in terms of both road holding and ride comfort at the human comfort frequency range (1-8\,Hz). Moreover, with the characterized system uncertainties, the proposed $\mu$-synthesis-based control scheme provides a more robust suspension performance enhancement of the PALS full car,
as compared to conventional $H_{\infty}$-based control schemes that significantly under-perform at deviated mass parameters.

The proposed multi-objective PID and $\mu$-synthesis control schemes are further merged to enable all functions of the PALS in practice, given selected uncertainties. As compared to the conventional hybrid control of multi-objective PID and $H_{\infty}$ control for nominal payload, the combined multi-objective PID $\mu$-synthesis control preserves the high frequency ride comfort and road holding improvement, without deteriorating the low frequency chassis leveling performance. In addition, when the payload is varied, the proposed multi-objective PID $\mu$-synthesis control significantly outperforms the conventional hybrid control scheme, especially at high frequencies. In terms of a comparison between the two proposed schemes of $\mu$-synthesis control and the multi-objective PID $\mu$-synthesis control, the latter is offering a similar performance as the former in almost all cases of driving maneuvers (at high frequencies), while at the same time, it can also effectively control low frequency dynamics that $\mu$-synthesis control does not cover. Numerical simulations of various ISO maneuvers, such as steady-state cornering, step steer, brake in turn, speed bump and random road C with a high fidelity model of the vehicle with PALS, have been provided to illustrate the effectiveness of the proposed methods.
\\
\appendix
\begin{table}[htb!]
\centering
\caption{Main Parameters of Original and PALS Retrofitted SUV full car, and linear equivalent model (F: Front, R: Rear)}\label{tab1-1}
\label{tab:vehicle data}
\scalebox{1}{\begin{tabular*}{1\columnwidth}{@{\extracolsep{\fill}}l @{\extracolsep{\fill}} c@{\extracolsep{\fill}}
c@{\extracolsep{\fill}}c @{\extracolsep{\fill}} c}
\hline
\hline
 Parameters & Value\\
 \hline
 \multicolumn{4}{c}{Original vehicle \cite{arana2016series,yuchassis}}\\
 \hline
 CMC Height & 0.71\,m \\
 F/R Weight distribution & 50/50\,$\%$ \\
 Total/Sprung mass & 2950/2700\,$kg$ \\
 F/R Suspension spring stiffness & 150/200\,$\frac{kN}{m}$ \\
 F/R Tire stiffness & 290\,$\frac{kN}{m}$ \\
 F/R Tire suspension damping & 300\,$\frac{Ns}{m}$ \\
 F/R Installation ratio & 0.58/0.50\\
 \hline
  \multicolumn{4}{c}{Linear equivalent model PALS full car }\\
  \hline
 F/R Nominal Suspension & 2355\, /\,2002\,$\frac{Ns}{m}$ \\
 damping ($\bar{c}_{{eq}_{f}}/\bar{c}_{{eq}_{r}}$) &   \\
 F/R Equivalent Suspension  & 53.5\,/\,$\frac{kN}{m}$ \\
 stiffness\,(${k_{eq}}_{f}/{k_{eq}}_{r}$) & \\
  F/R Tire stiffness\,(${k_{t}}_{f}/{k_{t}}_{r}$) & 290\,$\frac{kN}{m}$ \\
 F/R Tire damping\,(${c_{t}}_{f}/{c_{t}}_{r}$) & 300\,$\frac{Ns}{m}$ \\
 F/R Track width\,($t_f/t_r$) & 1.677\,/\,1.696\,$m$ \\
 F/R Wheelbase\,($a_f/a_r$) & 1.538\,/\,1.538\,$m$\\
 Nominal sprung mass\,($M_{\text{nom}}$) & 2700\,kg\\
 \hline
 \multicolumn{4}{c}{PALS retrofit (actuator per corner) \cite{arana2016series,yuchassis}}\\
 \hline
 F$\&$R Actuator mass & 12\,kg\\
 F$\&$R Gear Ratio & 66 \\
 F/R Low speed shaft (LSS) & 166/165\,$N\!\!\cdot\!m$ \\
 continuous torque & \\
 F/R LSS peak torque & 273/273\,$N\!\!\cdot\!m$ \\
\hline
\hline
\\
\end{tabular*}}
\end{table}

\begin{table}[htb!]
\centering
\caption{PID tuning parameters in `PALS-PID' `PALS-PID-$H_{\infty}$' and `PALS-PID-$\mu$' control schemes}\label{tab1-2}
\scalebox{1}{\begin{tabular*}{1\columnwidth}{@{\extracolsep{\fill}}l @{\extracolsep{\fill}} c@{\extracolsep{\fill}}
c@{\extracolsep{\fill}}c @{\extracolsep{\fill}} c@{\extracolsep{\fill}} c@{\extracolsep{\fill}}c}
\hline
\hline
Controller & Aim & Axle & P & I & D\\
 \hline
`PALS-PID' & pitch & F$\&$R & 1000 & 20000 & 4\\
 & roll & F$\&$R & 500 & 5000 & 4\\
`PALS-PID-$H_{\infty}$' & pitch & F$\&$R & 1000 & 5000 & 4\\
 & roll & F$\&$R & 500 & 2500 & 4\\
`PALS-PID-$\mu$'& pitch & F$\&$R & 1000 & 6000 & 4\\
& roll & F$\&$R & 2000 & 2500 & 4\\
\hline
\hline
\end{tabular*}}
\end{table}

\bibliographystyle{IEEEtran}
\bibliography{ref}

\begin{thebibliography}{10}
\providecommand{\url}[1]{#1}
\csname url@samestyle\endcsname
\providecommand{\newblock}{\relax}
\providecommand{\bibinfo}[2]{#2}
\providecommand{\BIBentrySTDinterwordspacing}{\spaceskip=0pt\relax}
\providecommand{\BIBentryALTinterwordstretchfactor}{4}
\providecommand{\BIBentryALTinterwordspacing}{\spaceskip=\fontdimen2\font plus
\BIBentryALTinterwordstretchfactor\fontdimen3\font minus
  \fontdimen4\font\relax}
\providecommand{\BIBforeignlanguage}[2]{{%
\expandafter\ifx\csname l@#1\endcsname\relax
\typeout{** WARNING: IEEEtran.bst: No hyphenation pattern has been}%
\typeout{** loaded for the language `#1'. Using the pattern for}%
\typeout{** the default language instead.}%
\else
\language=\csname l@#1\endcsname
\fi
#2}}
\providecommand{\BIBdecl}{\relax}
\BIBdecl

\bibitem{sharp1987road}
R.~Sharp and D.~Crolla, ``Road vehicle suspension system design-a review,''
  \emph{Vehicle system dynamics}, vol.~16, no.~3, pp. 167--192, 1987.

\bibitem{Aranaphdthesis}
C.~Arana, ``Active variable geometry suspension for cars,'' Ph.D. dissertation,
  Imperial College London, 2015.

\bibitem{talib2013self}
M.~H.~A. Talib and I.~Z.~M. Darns, ``Self-tuning pid controller for active
  suspension system with hydraulic actuator,'' in \emph{2013 IEEE Symposium on
  Computers \& Informatics (ISCI)}.\hskip 1em plus 0.5em minus 0.4em\relax
  IEEE, 2013, pp. 86--91.

\bibitem{ahmed2015pid}
A.~E.-N.~S. Ahmed, A.~S. Ali, N.~M. Ghazaly, and G.~Abd~el Jaber, ``Pid
  controller of active suspension system for a quarter car model,''
  \emph{International Journal of Advances in Engineering \& Technology},
  vol.~8, no.~6, p. 899, 2015.

\bibitem{khodadadi2018self}
H.~Khodadadi and H.~Ghadiri, ``Self-tuning pid controller design using fuzzy
  logic for half car active suspension system,'' \emph{International Journal of
  Dynamics and Control}, vol.~6, no.~1, pp. 224--232, 2018.

\bibitem{sun2012adaptive}
W.~Sun, H.~Gao, and O.~Kaynak, ``Adaptive backstepping control for active
  suspension systems with hard constraints,'' \emph{IEEE/ASME transactions on
  mechatronics}, vol.~18, no.~3, pp. 1072--1079, 2012.

\bibitem{pang2019adaptive}
H.~Pang, X.~Zhang, and Z.~Xu, ``Adaptive backstepping-based tracking control
  design for nonlinear active suspension system with parameter uncertainties
  and safety constraints,'' \emph{ISA transactions}, vol.~88, pp. 23--36, 2019.

\bibitem{deshpande2014disturbance}
V.~S. Deshpande, B.~Mohan, P.~Shendge, and S.~Phadke, ``Disturbance observer
  based sliding mode control of active suspension systems,'' \emph{Journal of
  Sound and Vibration}, vol. 333, no.~11, pp. 2281--2296, 2014.

\bibitem{liu2020adaptive}
Y.-J. Liu and H.~Chen, ``Adaptive sliding mode control for uncertain active
  suspension systems with prescribed performance,'' \emph{IEEE Transactions on
  Systems, Man, and Cybernetics: Systems}, 2020.

\bibitem{wang2017robust}
G.~Wang, C.~Chen, and S.~Yu, ``Robust non-fragile finite-frequency $h_{\infty}$
  static output-feedback control for active suspension systems,''
  \emph{Mechanical Systems and Signal Processing}, vol.~91, pp. 41--56, 2017.

\bibitem{jing2014output}
H.~Jing, X.~Li, and H.~Karimi, ``Output-feedback based $h_{\infty}$ control for
  active suspension systems with control delay,'' \emph{IEEE Transactions on
  Industrial Electronics}, vol.~61, no.~1, pp. 436--446, 2014.

\bibitem{gohrle2012active}
C.~G{\"o}hrle, A.~Wagner, A.~Schindler, and O.~Sawodny, ``Active suspension
  controller using mpc based on a full-car model with preview information,'' in
  \emph{2012 American Control Conference (ACC)}.\hskip 1em plus 0.5em minus
  0.4em\relax IEEE, 2012, pp. 497--502.

\bibitem{gohrle2013design}
C.~Gohrle, A.~Schindler, A.~Wagner, and O.~Sawodny, ``Design and vehicle
  implementation of preview active suspension controllers,'' \emph{IEEE
  Transactions on Control Systems Technology}, vol.~22, no.~3, pp. 1135--1142,
  2013.

\bibitem{wang2012hierarchical}
W.-Y. Wang, M.-C. Chen, and S.-F. Su, ``Hierarchical t--s fuzzy-neural control
  of anti-lock braking system and active suspension in a vehicle,''
  \emph{Automatica}, vol.~48, no.~8, pp. 1698--1706, 2012.

\bibitem{sun2014vibration}
W.~Sun, H.~Gao, and O.~Kaynak, ``Vibration isolation for active suspensions
  with performance constraints and actuator saturation,'' \emph{IEEE/ASME
  transactions on mechatronics}, vol.~20, no.~2, pp. 675--683, 2014.

\bibitem{kim2011height}
H.~Kim and H.~Lee, ``Height and leveling control of automotive air suspension
  system using sliding mode approach,'' \emph{IEEE Transactions on Vehicular
  Technology}, vol.~60, no.~5, pp. 2027--2041, 2011.

\bibitem{kim2011fault}
H.~Kim and H.~Lee, ``Fault-tolerant control algorithm for a four-corner
  closed-loop air suspension system,'' \emph{IEEE Transactions on industrial
  Electronics}, vol.~58, no.~10, pp. 4866--4879, 2011.

\bibitem{yu2018parallel}
M.~Yu, C.~Arana, S.~A. Evangelou, D.~Dini, and G.~D. Cleaver, ``Parallel active
  link suspension: A quarter-car experimental study,'' \emph{IEEE/ASME
  Transactions on Mechatronics}, vol.~23, no.~5, pp. 2066--2077, 2018.

\bibitem{yuchassis}
M.~Yu, S.~A. Evangelou, and D.~Dini, ``Chassis leveling control with parallel
  active link suspension,'' in \emph{14th International Symposium on Advanced
  Vehicle Control (AVEC)}, 2018.

\bibitem{yu2018control}
M.~Yu, S.~A. Evangelou, and D.~Dini, ``Control design for a quarter car test
  rig with parallel active link suspension,'' in \emph{2018 Annual American
  Control Conference (ACC)}.\hskip 1em plus 0.5em minus 0.4em\relax IEEE, 2018,
  pp. 3227--3232.

\bibitem{zhou1998essentials}
K.~Zhou and J.~C. Doyle, \emph{Essentials of robust control}.\hskip 1em plus
  0.5em minus 0.4em\relax Prentice hall Upper Saddle River, NJ, 1998, vol. 104.

\bibitem{feng2020uncertainties}
Z.~Feng, M.~Yu, C.~Cheng, S.~A. Evangelou, I.~M. Jaimoukha, and D.~Dini,
  ``Uncertainties investigation and $\mu$-synthesis control design for a full
  car with series active variable geometry suspension,''
  \emph{IFAC-PapersOnLine}, vol.~53, no.~2, pp. 13\,882--13\,889, 2020.

\bibitem{arana2014series}
C.~Arana, S.~A. Evangelou, and D.~Dini, ``Series active variable geometry
  suspension for road vehicles,'' \emph{IEEE/ASME Transactions On
  Mechatronics}, vol.~20, no.~1, pp. 361--372, 2014.

\bibitem{yu2021series}
M.~Yu, C.~Cheng, S.~A. Evangelou, and D.~Dini, ``Series active variable
  geometry suspension: Full-car prototyping and road testing,'' \emph{IEEE/ASME
  Transactions on Mechatronics}, 2021.

\bibitem{yu2019position}
M.~Yu, S.~A. Evangelou, and D.~Dini, ``Position control of parallel active link
  suspension with backlash,'' \emph{IEEE Transactions on Industrial
  Electronics}, vol.~67, no.~6, pp. 4741--4751, 2019.

\bibitem{chen2014mu}
Z.~Chen, B.~Yao, and Q.~Wang, ``mu-synthesis-based adaptive robust control of
  linear motor driven stages with high-frequency dynamics: A case study,''
  \emph{IEEE/ASME Transactions on Mechatronics}, vol.~20, no.~3, pp.
  1482--1490, 2014.

\bibitem{kaleemullah2014active}
M.~Kaleemullah and W.~F. Faris, ``Active suspension control of vehicle with
  uncertainties using robust controllers,'' \emph{International Journal of
  Vehicle Systems Modelling and Testing}, vol.~9, no. 3-4, pp. 293--310, 2014.

\bibitem{ISO_2631-1:1997}
ISO, ``2631-1:1997, {M}echanical vibration and shock - {E}valuation of human
  exposure to whole-body vibration - part 1: General requirements,'' Geneva,
  Switzerland, 1997.

\bibitem{ISO_8608-2016}
ISO, ``8608:2016, {M}echanical vibration - {R}oad surface profiles - reporting
  of measured data,'' Geneva, Switzerland, 2016.

\bibitem{iso20124138}
ISO, ``4138: Passenger cars--steady-state circular driving behaviour--open-loop
  test methods,'' \emph{ISO: Geneva, Switzerland}, 2012.

\bibitem{iso2011road}
ISO, ``7401:2011,road vehicles--lateral transient response test
  methods--open-loop test methods,'' 2011.

\bibitem{ISO_7975-2006}
ISO, ``7975:2006,passenger cars – braking in a turn – open loop test
  procedure,'' 2006.

\bibitem{arana2016series}
C.~Arana, S.~A. Evangelou, and D.~Dini, ``Series active variable geometry
  suspension application to chassis attitude control,'' \emph{IEEE/ASME
  Transactions on Mechatronics}, vol.~21, no.~1, pp. 518--530, 2016.

\end{thebibliography}

\end{document}